Technical Report

# Evaluation of Contactless Smartcard Antennas

Michael Roland

University of Applied Sciences Upper Austria
Josef Ressel Center u'smile
michael.roland@fh-hagenberg.at

Michael Hölzl

Johannes Kepler University Linz
Institute of Networks and Security
hoelzl@ins.jku.at

**Abstract** This report summarizes the results of our evaluation of antennas of contactless and dual interface smartcards and our ideas for user-switchable NFC antennas. We show how to disassemble smartcards with contactless capabilities in order to obtain the bare chip module and the bare antenna wire. We examine the design of various smartcard antennas and present concepts to render the contactless interface unusable. Finally, we present ideas and practical experiments to make the contactless interface switchable by the end-user.

This work has been carried out within the scope of "u'smile", the Josef Ressel Center for User-Friendly Secure Mobile Environments, funded by the Christian Doppler Gesellschaft, A1 Telekom Austria AG, Drei-Banken-EDV GmbH, LG Nexera Business Solutions AG, NXP Semiconductors Austria GmbH, and Österreichische Staatsdruckerei GmbH. Moreover, this work has been carried out in close cooperation with the project "High Speed RFID" within the EU programme "Regionale Wettbewerbsfähigkeit OÖ 2007–2013 (Regio 13)" funded by the European Regional Development Fund (ERDF) and the Province of Upper Austria (Land Oberösterreich).

UNIVERSITY OF APPLIED SCIENCES UPPER AUSTRIA

# Contents

# 1. Introduction

The recent addition of contactless capabilities to the Austrian “Bankomatkarte” (as well as to most new credit cards issued by banks all across Europe and the rest of the world) raised many security and privacy concerns. While these concerns are often hyped in the media, contactless smartcard systems (particularly payment cards) employ various mechanisms to mitigate attacks on the protocol layer as well as on an organizational layer (e.g. automated fraud detection systems). Nevertheless, the contactless interface is an additional attack vector that leads to certain new attack scenarios (cf. [1, 4, 6, 10, 14]). Particularly the fact that a certain number of low-value contactless transactions can be performed without the need for a PIN code lead to an outcry in the media.

This report aims at giving an overview of the construction of such smartcards with contactless capabilities. We start by providing an insight of what is hidden below the plastic surface of these smartcards, and by explaining how contactless and dual interface smartcards could be disassembled in order to get access to the bare chip module and the bare antenna wire. We further analyze the construction principle and the antenna design of various dual interface smartcards. Based on this analysis, we outline and evaluate ideas for rendering the contactless functionality of these smartcards permanently unusable without affecting any other functionality. Finally, we sketch concepts and ideas for improving the security of smartcards that expose contactless capabilities by adding a notion of explicit user-consent that is required for accessing contactless functionality. We partially verify the technical feasibility of these concepts with prototypical implementations of switchable smartcard antennas.

# 2. Disassembling Smartcards

For our evaluation of switchable NFC antennas we first needed to get smartcard chips that we could later attach to our customized antenna designs. Moreover, we wanted to analyze the layout of antennas in existing plastic smartcards. Therefore, we tried to disassemble some cards.

## 2.1 Construction Principle of a Plastic Smartcard

This section shows an example for one construction principle of a typical smartcard. A comprehensive description of different construction principles of plastic chip cards can be found in [13].

A smartcard typically consists of several layers (see Fig. 1) of plastic foils that are laminated (besides lamination of foils, other methods to construct the card body exist) on top of each other. These layers could, for instance, be

1. a transparent front cover sheet,
2. a sheet with printed graphical elements that should be visible on the front of the card,
3. a plastic card body sheet with copper wire laid into it as the antenna,
4. a sheet with printed graphical elements that should be visible on the back of the card, and
5. a transparent back cover sheet.

Several variations of this (in terms of the number and type of layers, embedding of the antenna, etc.) are possible. For instance,

- a layer containing a magnetic stripe could be added,
- the cover sheets could be made printable after card manufacturing using a thermo transfer printer or could contain a signature field, a hologram, etc., or
- the antenna could be printed on (or etched from) an inlay foil that is laminated between two card body sheets.

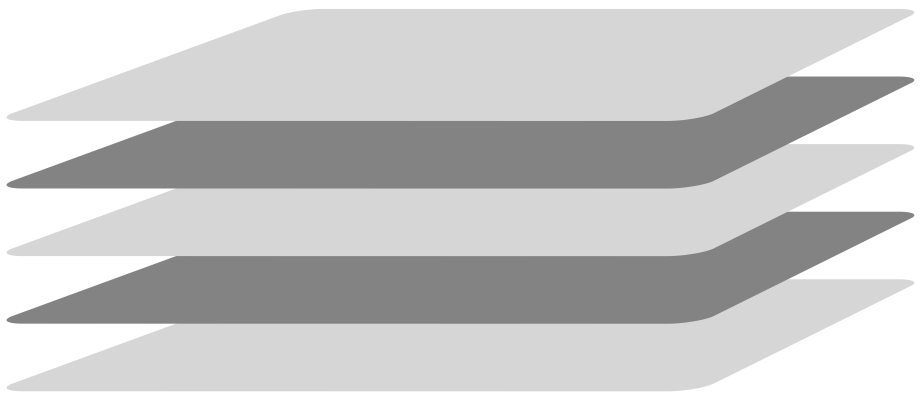

Figure 1: Stacked layers of a plastic card

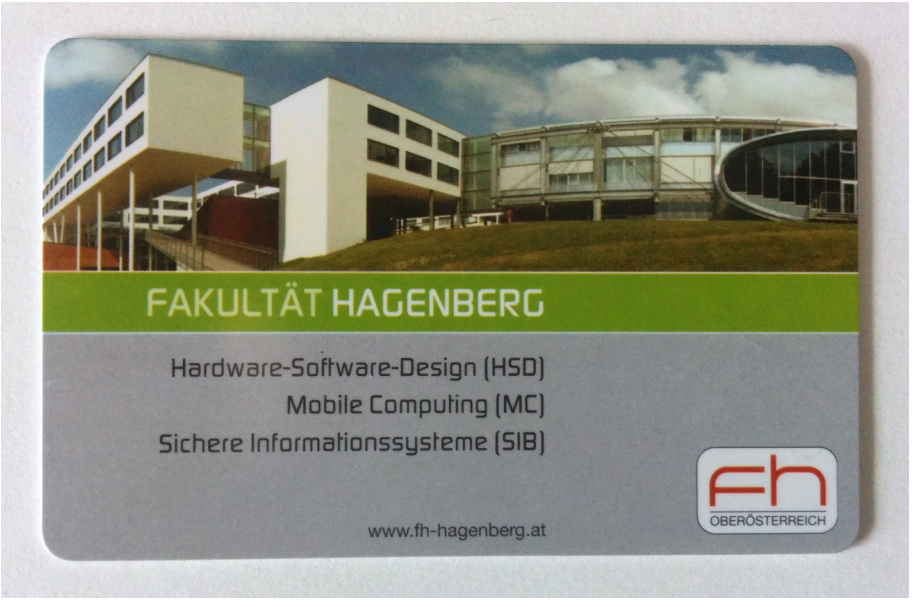

(a) Laminated plastic card

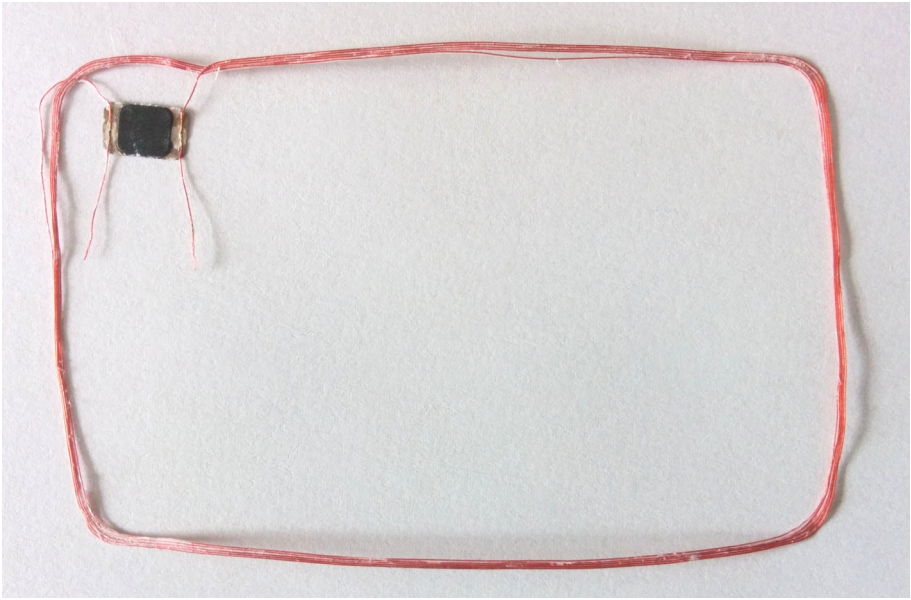

(b) Antenna with chip from a dissolved plastic card

Figure 2: MIFARE Classic card in ID-1 format

Chip modules are either added before lamination (contactless-only cards) by gluing, welding or soldering them to the antenna or after lamination (contact + contactless dual interface cards) by gluing them into a hole that was cut out of the card body with a mill.

## 2.2 Dissolving a MIFARE Classic Card

We found that our MIFARE Classic cards can be dissolved using acetone or paint thinner ("Nitroverdünnung"). After putting a card into a jar filled with paint thinner and letting it rest in there for some days, the plastic layers of the card fall apart. The inner layer (or layers?) soak with thinner and expand to a sticky soft mass, while the thin outer layers do not soak but get ripped into thin stripes as they stick to the expanding inner layers. The paint used for the printed graphical elements dissolves in the thinner.

Then, the chip and the attached antenna can be ripped out of the remains of the card body. The card (before dissolving) and the bare antenna and chip module (after dissolving) are shown in Fig. 2. With this model, the chip is welded to the antenna. The antenna is made of thin copper wire (5 windings, approx. 77 mm × 47 mm) and was laid into one of the layers of the card body.

## 2.3 Extracting the Chip from a Dual Interface Smartcard

In order to get a bare dual interface smartcard chip module, we decided to disassemble a credit card (Fig. 3). To remove the chip module from the dual interface card, we heated the area around the chip module to a temperature of about 200 degrees Celsius using a heat gun. As a result, the adhesive bond that was used to mount the module to the card body disjoint and the chip module could be removed from the card body using tweezers. Fig. 4 shows the front and the back of the extracted chip module and the milled hole in the card body where the chip module was attached. On the left and the right side of the hole there are the contact pads that are used to attach the antenna to the chip module.

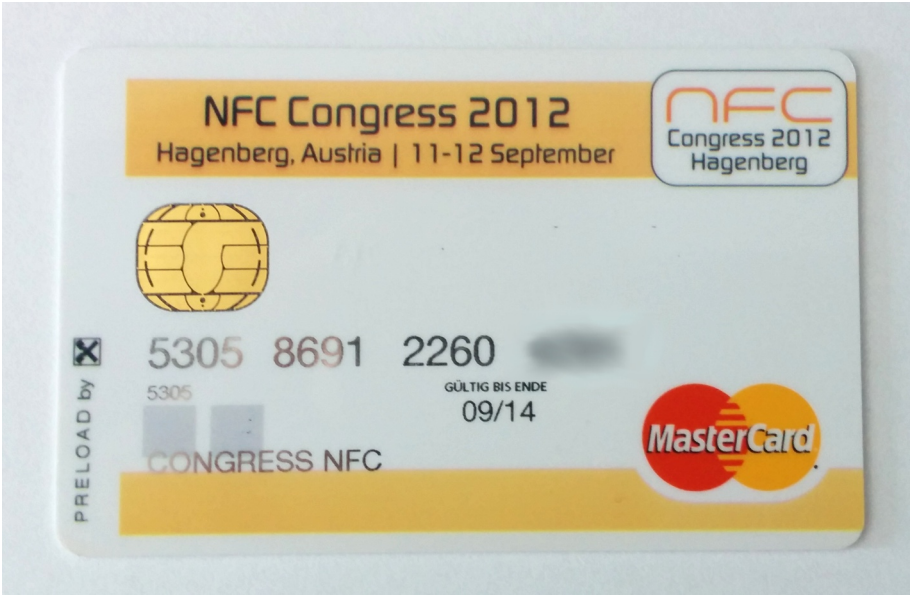

Figure 3: Dual interface smartcard in ID-1 format

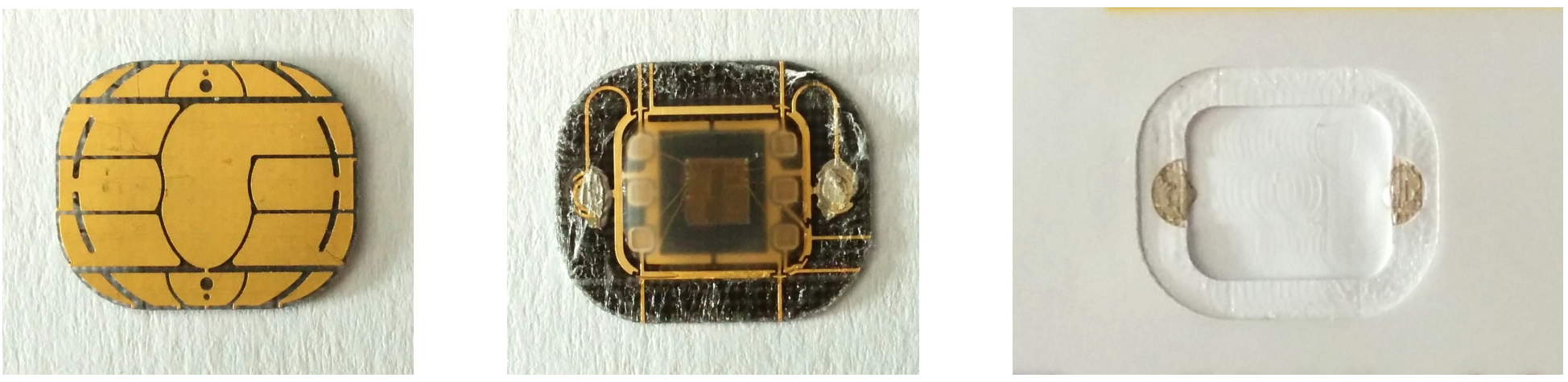

(a) front of module (b) back of module (c) hole in card body

Figure 4: Dual interface chip module and the hole in the plastic card body where the chip module was attached

This method of extracting the chip module has destructive effects to the card body. The card body gets soft and starts to melt at this temperature. Consequently, the card deforms and slightly shrinks at the heated area (see Fig. 5).

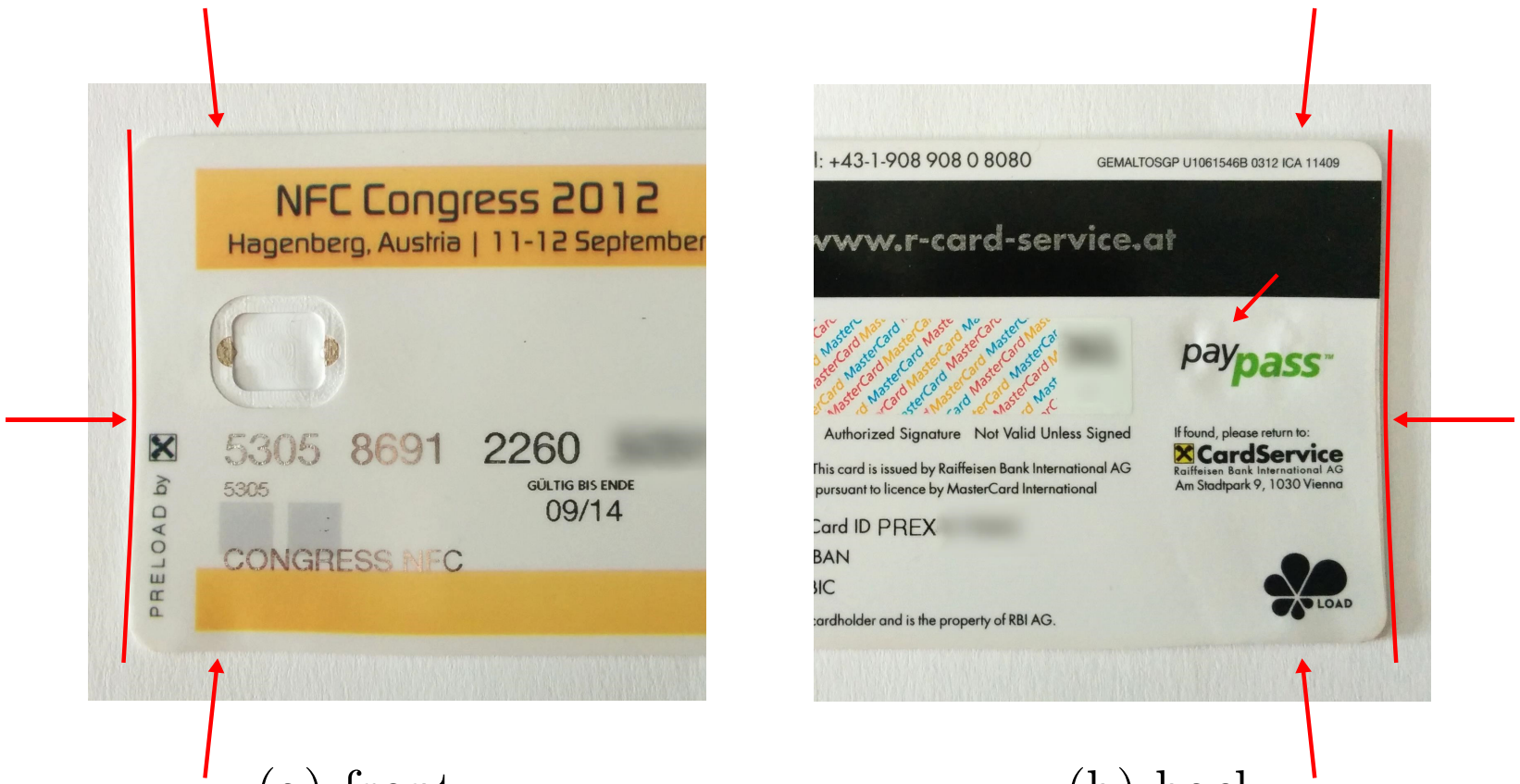

(a) front (b) back

Figure 5: Deformation of the plastic card after extracting the chip module

# 3. Analysis of Dual Interface Smartcard Antennas

## 3.1 Non-Destructive Analysis

As a next step we wanted to examine the shape and size of typical smartcard antennas without destroying the cards. The best way to perform such an analysis would possibly be to X-ray the cards. However, as we did not have access to an X-ray machine, we used a bright flashlight. The light of the flashlight shines through the card and is blocked by the antenna wire, the metal contacts of the chip module, and, to a significant part, also by the magnetic stripe. Hence, it is possible to see the silhouette of the antenna. While darker colors in the artwork make it more difficult to identify the shape of the antenna, for some cards with blank card body, we were even able to recognize the reddish-orange color of the enameled copper wire.

Besides a visual inspection of the antenna shape, we also measured the resonant frequencies based on the reflection coefficient (S11) observed with a vector network analyzer (Rohde & Schwarz FSH4) using an ID-1 sized single-loop probe that is coupled to the card-under-test.

## 3.2 Examination of Card Antennas

Using this method, we were able to measure the size, the position and the routing of the antenna wire for several cards. Table 1 gives an overview of the examined cards.

Table 1: Examined cards

| Manufacturer | Card type | Product identification or visible markings | |
|---|---|---|---|
| Austria Card | Austrian “Bankomatkarte” (Maestro) | 01/13 AUSTRIACARD 204/015 | Fig. 6 |
| Austria Card | Visa | 09/14 AUSTRIACARD 54833/004 | Fig. 7 |
| Winter AG / Trüb AG | MasterCard picture card | ICA 7751 Winter/Trüb 02/11 19893500 | Fig. 8 |
| Gemalto | MasterCard picture card | GEMALTOSGP U1061546B 0312 ICA 11409 | Fig. 9 |
| Gemalto | blank card | IDCore 3010 | Fig. 10 |
| unknown | blank card | Athena IDProtect | Fig. 11 |
| unknown | blank card | NXP JCOP41 V2.3.1 | Fig. 12(a) |
| unknown | blank card | NXP J3A081 DI / JCOP V2.4.1 R2 | Fig. 12(b) |
| unknown | blank card | NXP J3D081 DI / JCOP V2.4.2 R2 | Fig. 12(c) |
| unknown | blank card | NXP JCOP41 engineering sample | Fig. 12(d) |

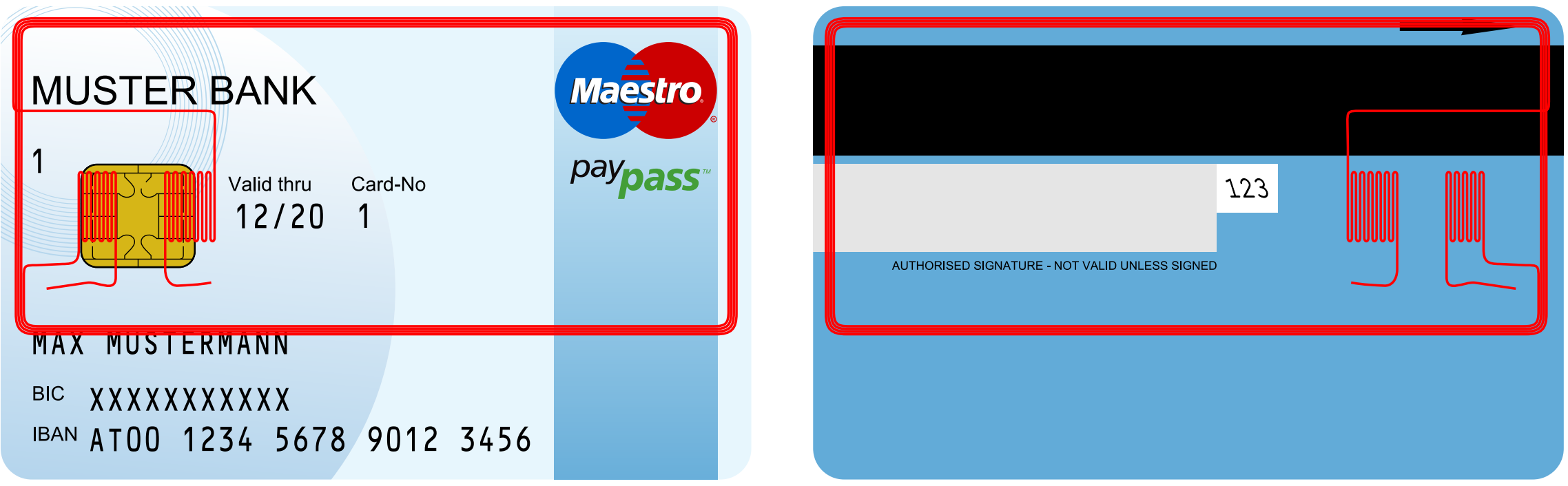

(a) front (b) back

Figure 6: Maestro card (Austrian “Bankomatkarte”) manufactured by Austria Card

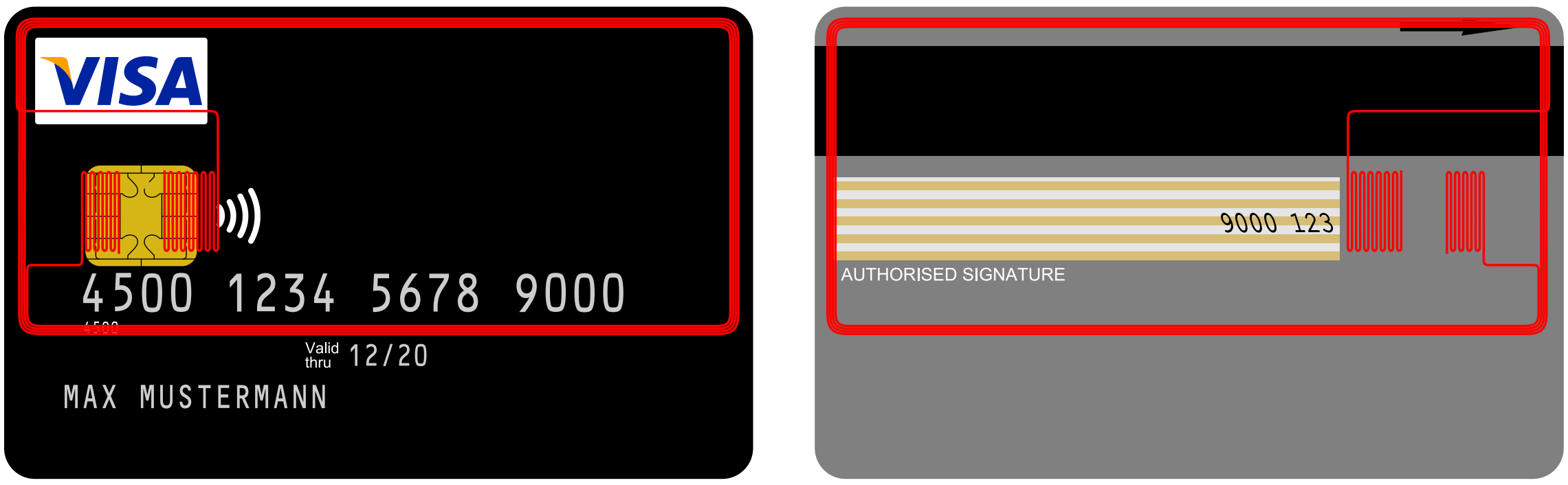

(a) front (b) back

Figure 7: Visa card manufactured by Austria Card

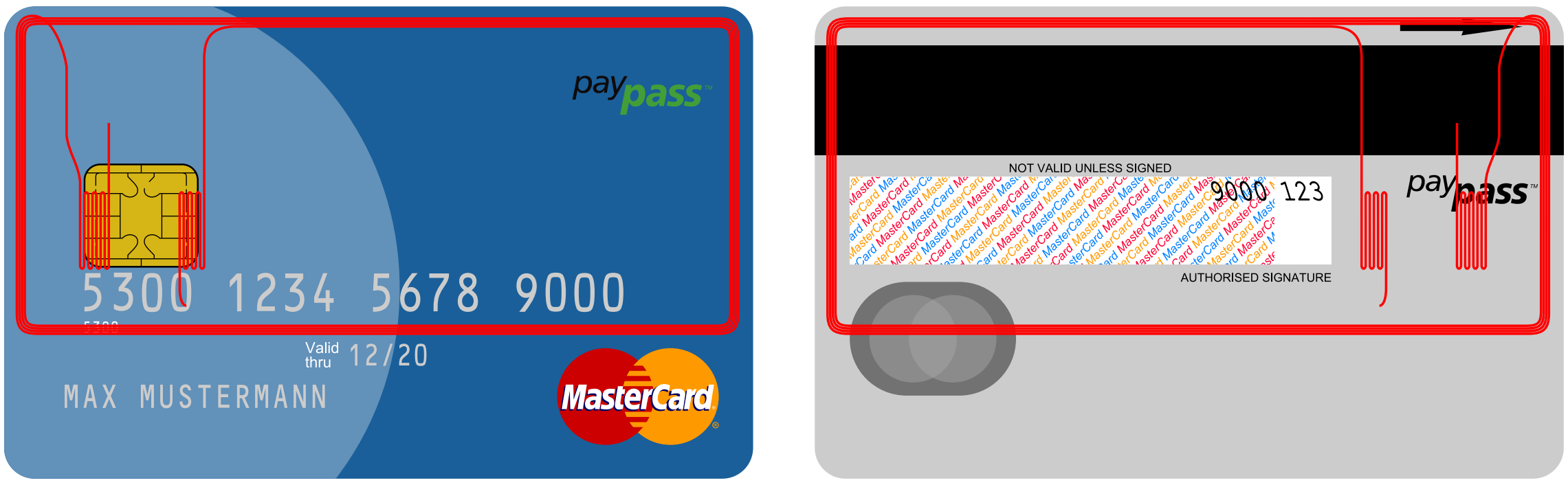

(a) front (b) back

Figure 8: MasterCard manufactured by Winter AG / Trüb AG

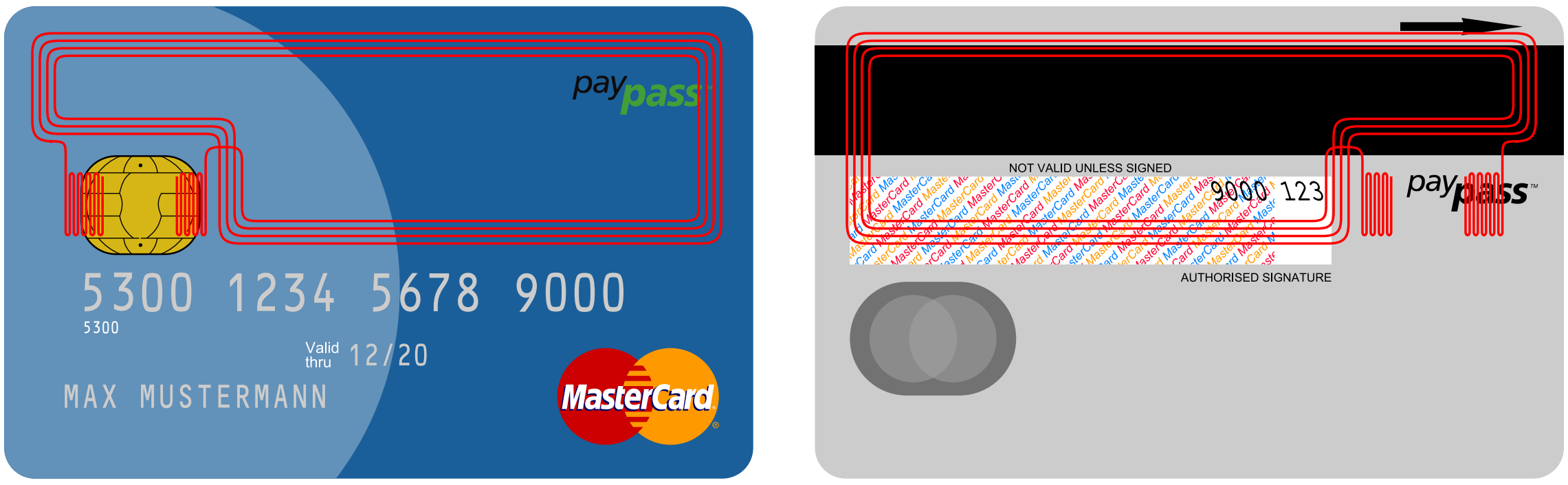

(a) front (b) back

Figure 9: MasterCard manufactured by Gemalto

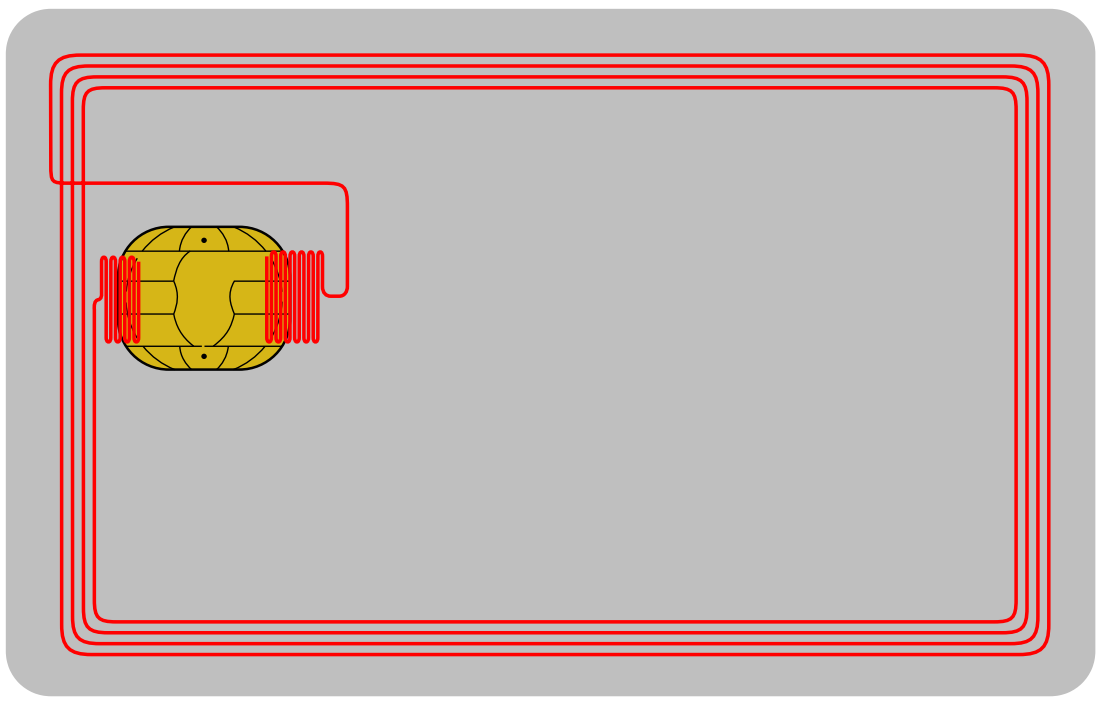

Figure 10: Gemalto IDCore 3010 manufactured by Gemalto

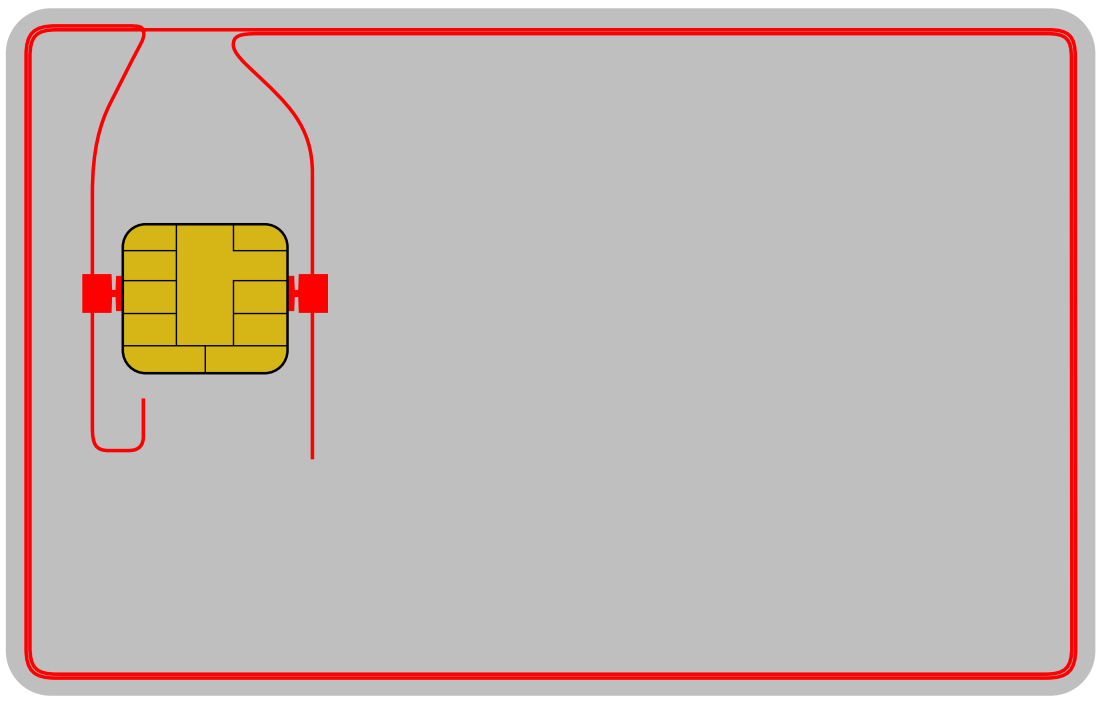

Figure 11: Athena IDProtect card (unknown manufacturer)

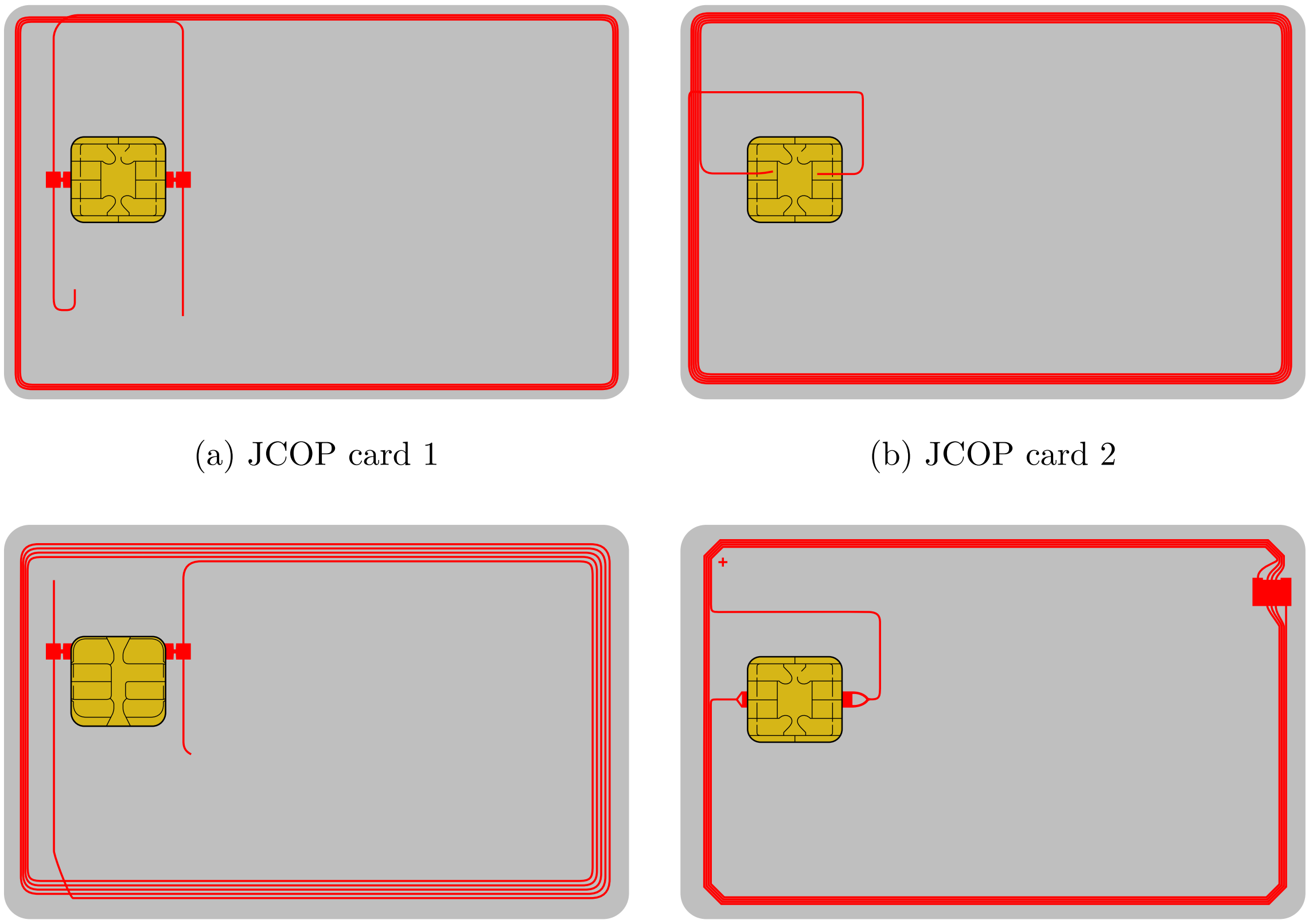

(a) JCOP card 1

(b) JCOP card 2

(c) JCOP card 3

(d) JCOP card 4

Figure 12: Various JCOP cards (unknown manufacturers)

### 3.2.1 Manufacturing Process

We found that for most of the analyzed cards, the antenna is made of enameled wire (possibly copper wire, though we could not verify this with our non-destructive analysis method) and was laid (melted) into the card body. Typical indications for this are overlappings of the antenna wire (which require the wire to be insulated), curved wire endings (from the movement of the laying machine), and meander-shaped areas of wire around the antenna contacts of the chip module.

The meander-shaped (see Fig. 6, 7, 8, 9, and 10) endings allow for production variations of the antenna position with respect to the chip module, as they create a bigger area where the antenna contacts of the chip module can be attached to the antenna wire with conductive adhesive.

Some cards (see Fig. 11, 12(a), 12(c)) have rectangular contact pads that protrude beyond the area of the contact pads and connect the antenna wires with the chip module. For one card (see Fig. 12(b)), the endings of the antenna wire were laid straight horizontally below the chip module.

Table 2: Antenna geometries

| Card | (↔) Width | (↕) Height | Windings | Size in relation to card body |
|---|---|---|---|---|
| Fig. 6 | 80 mm | 34 mm | 4 | 2/3 of card body |
| Fig. 7 | 80 mm | 34 mm | 4 | 2/3 of card body |
| Fig. 8 | 80 mm | 34 mm | 4 | 2/3 of card body |
| Fig. 9 | 74 mm | 22 mm | 4 | 1/2 of card body |
| Fig. 10 | 74 mm | 44 mm | 4 | full card body |
| Fig. 11 | 80 mm | 49 mm | 2 | full card body |
| Fig. 12(a) | 80 mm | 49 mm | 3 | full card body |
| Fig. 12(b) | 80 mm | 49 mm | 5 | full card body |
| Fig. 12(c) | 78 mm | 46 mm | 5 | full card body |
| Fig. 12(d) | 78 mm | 48 mm | 4 | full card body |

Only one card that we analyzed (see Fig. 12(d)) had an etched antenna inlay. Clear indications for this are the plus marking in one corner of the inlay and the rectangular area in the upper-right corner that creates a bridge from the outer winding to the inside of the antenna.

#### 3.2.2 Antenna Geometry

We found variations in the antenna size and in the number of windings (see Table 2). For all cards, the antenna is about 80 millimeters wide (center of the antenna wires). For our blank cards, the windings span across the whole card and the antenna is about 50 millimeters tall. For the examined credit and debit cards, we found that the antenna is located either in the upper half (approx. 22 mm tall) or the upper two-thirds (approx. 34 mm tall) of the card body. That way, the antenna wires are located either directly above or directly below the embossed card number.

The number of windings ranges from two to five. All inspected credit and debit cards have 4 windings. For the blank cards, we found antenna designs with 2, 3, 4, and 5 windings.

#### 3.2.3 Resonant Frequency

We found huge variations in the resonant frequency of the tested smartcards (see Fig. 13 for all measurements). Resonant frequencies ranged from 14.5 to 76.5 MHz. The majority of cards had a resonant frequency around 18 MHz. Cards tuned to resonant frequencies between 14 and 18 MHz were also found to work best in combination with mobile phones. This behavior was expected as resonant frequencies close to the operating frequency (13.56 MHz) improve the energy transfer between the reader and the card [2].

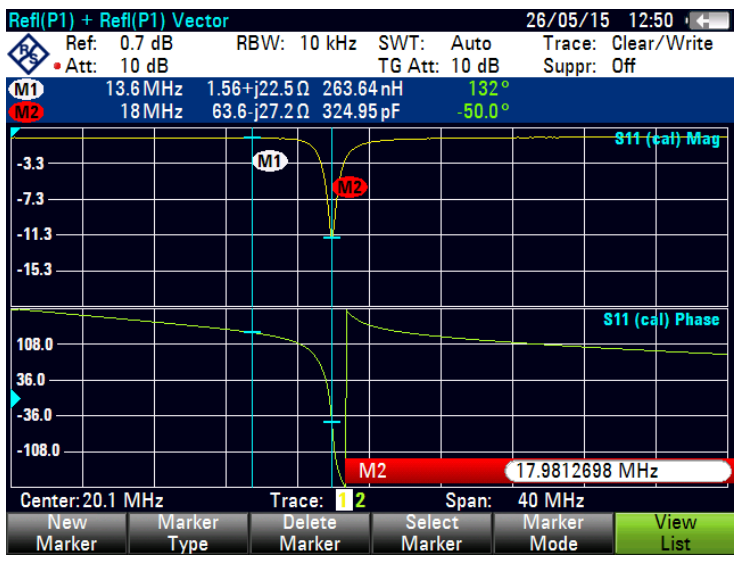

(a) Austria Card / Maestro, see Fig. 6

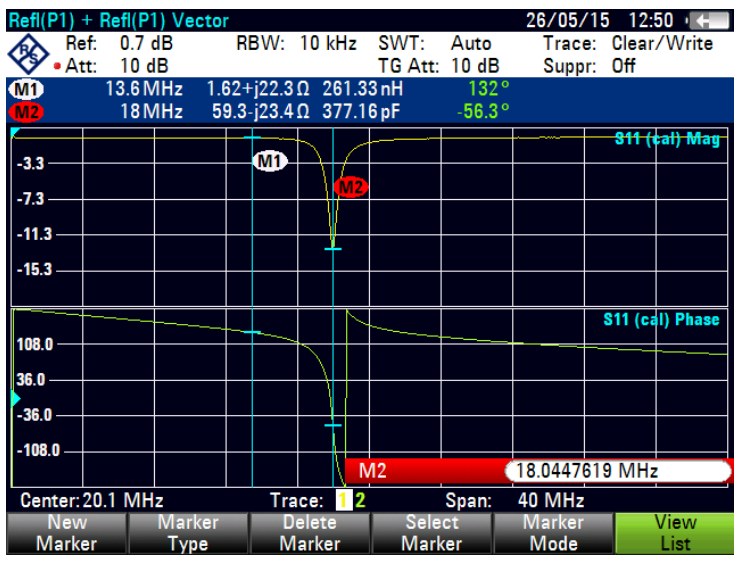

(b) Austria Card / Visa, see Fig. 7

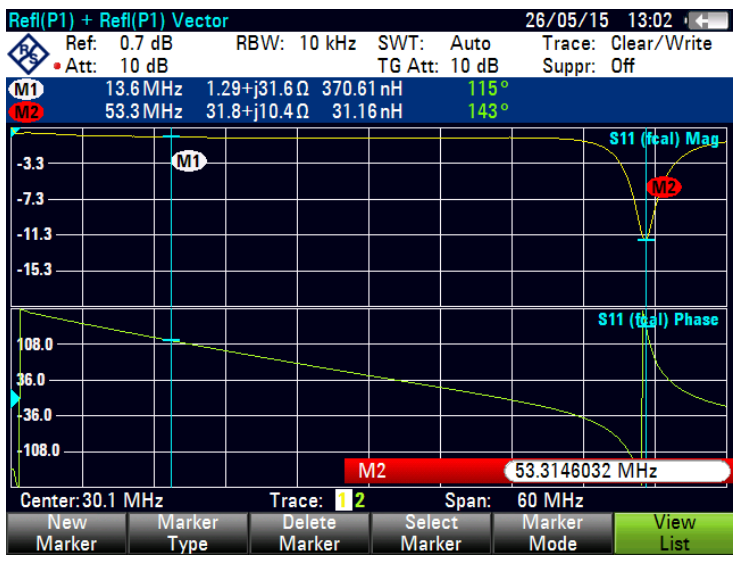

(c) Winter AG/Trüb AG / MasterCard, see Fig. 8

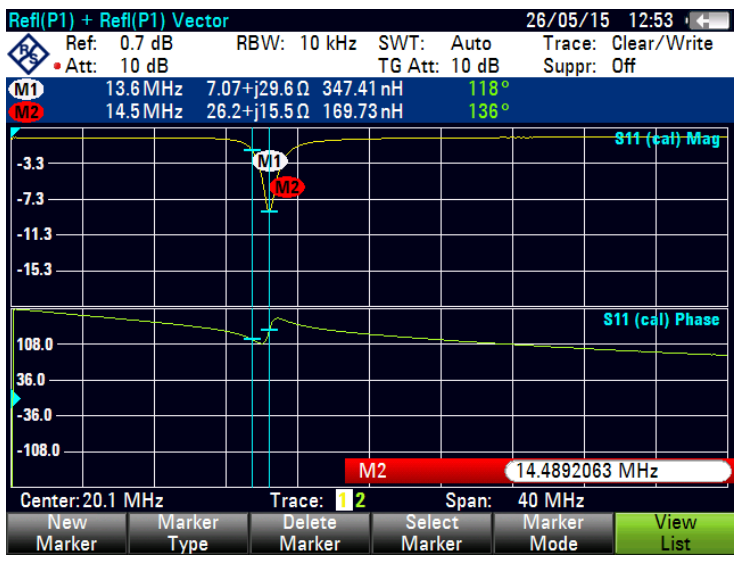

(d) Gemalto / MasterCard, see Fig. 9

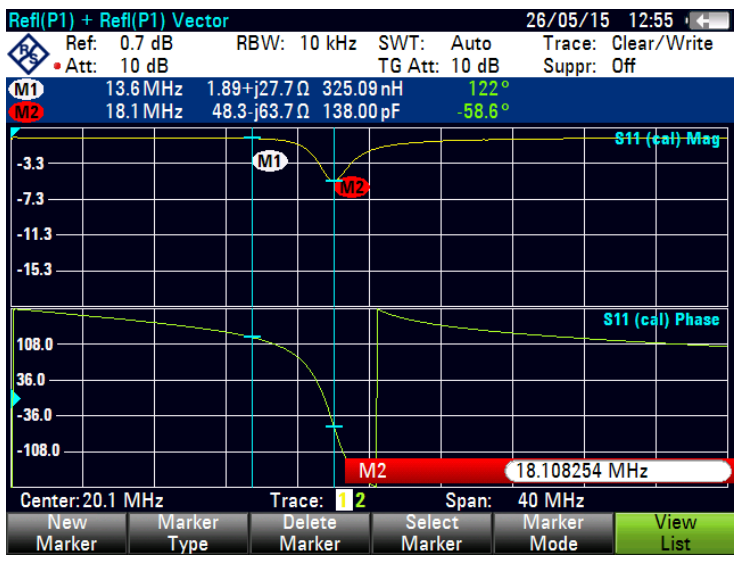

(e) Gemalto / IDCore 3010, see Fig. 10

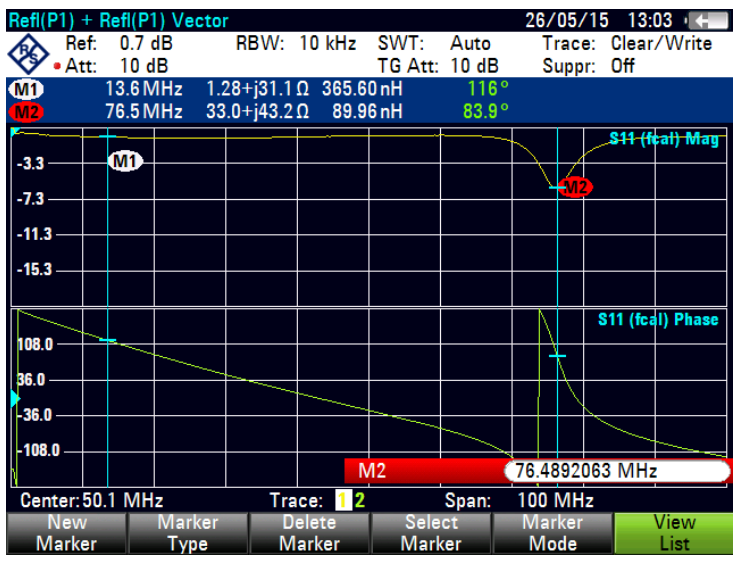

(f) Athena IDProtect, see Fig. 11

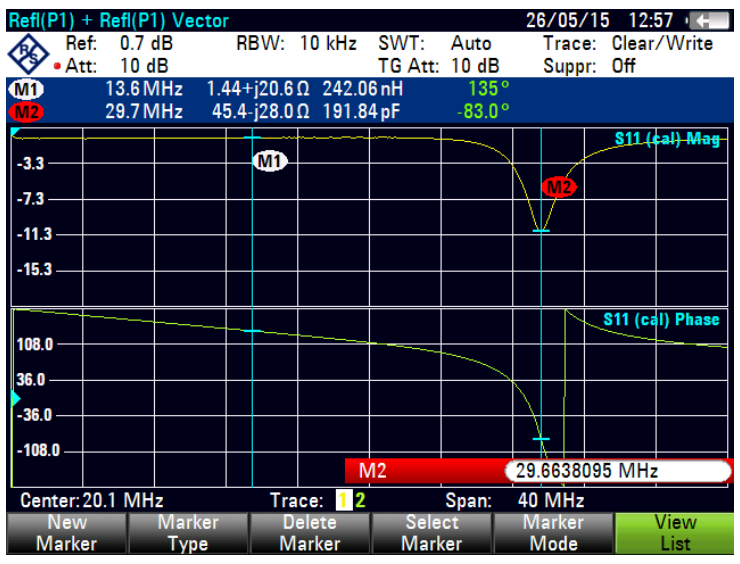

(g) Blank JCOP card 1, see Fig. 12(a)

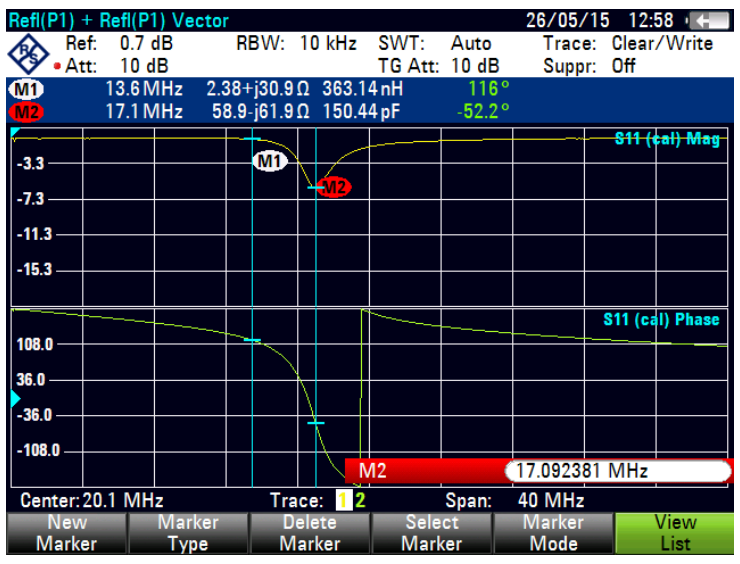

(h) Blank JCOP card 2, see Fig. 12(b)

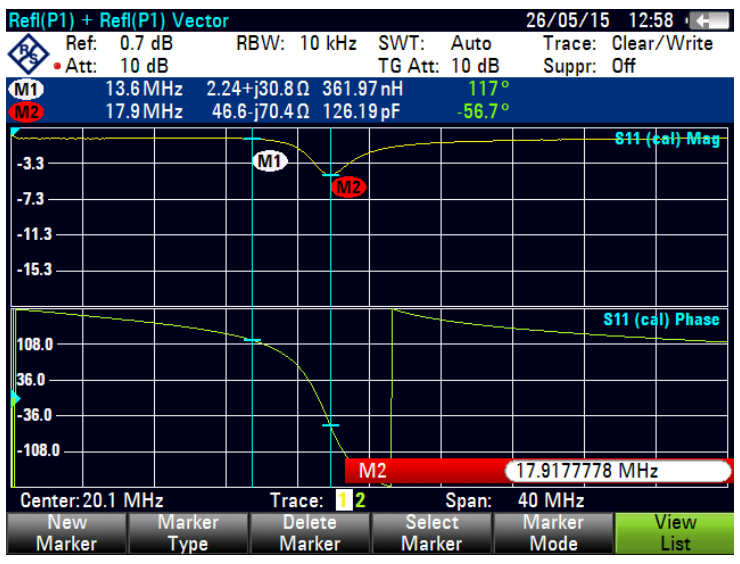

(i) Blank JCOP card 3, see Fig. 12(c)

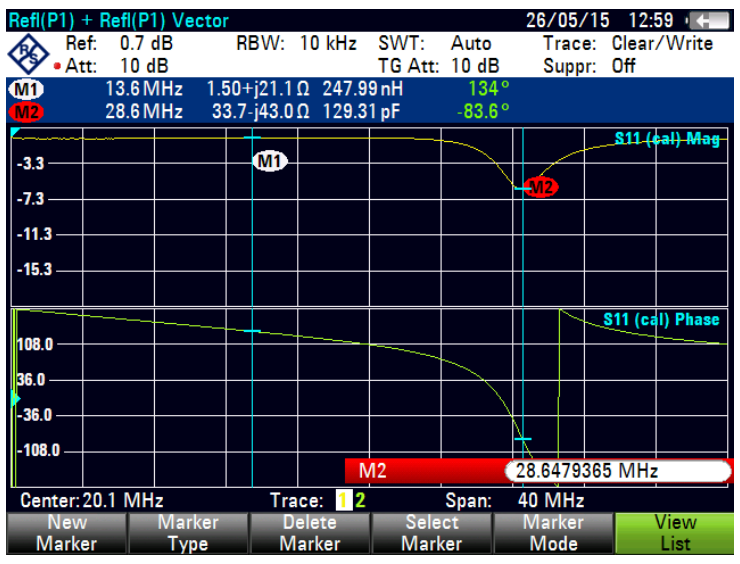

(j) Blank JCOP card 4, see Fig. 12(d)

| | Resonant frequency |
|---|---|
| (a) | 17.98 MHz |
| (b) | 18.04 MHz |
| (c) | 53.31 MHz |
| (d) | 14.49 MHz |
| (e) | 18.11 MHz |
| (f) | 76.49 MHz |
| (g) | 29.66 MHz |
| (h) | 17.09 MHz |
| (i) | 17.92 MHz |
| (j) | 28.65 MHz |

Figure 13: Resonant frequencies

# 4. Disabling the Contactless Interface of Dual Interface Cards

As both, the contact and the contactless functionality of a dual interface smartcard share the same chip (and, in fact, are just two different interfaces to the same smartcard logic), the only methods to disable the contactless interface while keeping the contact interface intact seem to be

1. to disable the contactless interface on a software level, or
2. to “unplug” the antenna from the chip.

As smartcards are security critical devices and changes to the software configuration are typically only possible using tightly controlled keys (or even only during certain manufacturing stages), the only option for an end-user would be to physically detach the antenna from the chip.

## 4.1 Cutting the Antenna Wire

Since we now know the location of the antenna within the card body, we wanted to check if it is possible to disable the contactless interface without influencing the remaining functionality of the card.

Looking at the cards in Fig. 6, 7, 8, and 9, the most promising location to cut the antenna is between the magnetic stripe and the signature field on the left edge of the card (when looking at the back of the card; right edge, when looking at the front of the card). We chose this position as it will not influence the magnetic stripe, is far away from the chip module, does not touch any of the visible security features (signature field, hologram, embossing), and does not make any text elements unreadable. Hence, the cut should not do any harm to the remaining functionality of the card (magnetic stripe, contact interface of the chip) and should not raise too much suspicion when using the card for payment. Moreover, this position is far away from the edge of the card that is inserted first into an ATM. Consequently, it will not trigger the mechanical shutter that card readers of many ATMs are equipped with in order to protect against insertion of anything that is not a bank card (cf. [19]).

Using a junior hacksaw, we cut approx. 10 mm deep into the card. We tested this with the MasterCard picture card manufactured by Gemalto (Fig. 9) as we had a couple of expired credit cards of this type. Figure 14 shows the position of the slit that we cut into the card body. Due to the chosen blade, the cut became approx. 1 mm wide. Cutting 10 mm deep assures that we cut through all windings of the antenna (and additionally allows for variations in the antenna position). Moreover, this type of cut would also work for the other payment card antennas as seen in Fig. 6, 7, and 8.

To get an in-depth insight on the effects of cutting the antenna windings, we cut the antenna wire by wire (from the outer loop to the inner loop) and analyzed the effects

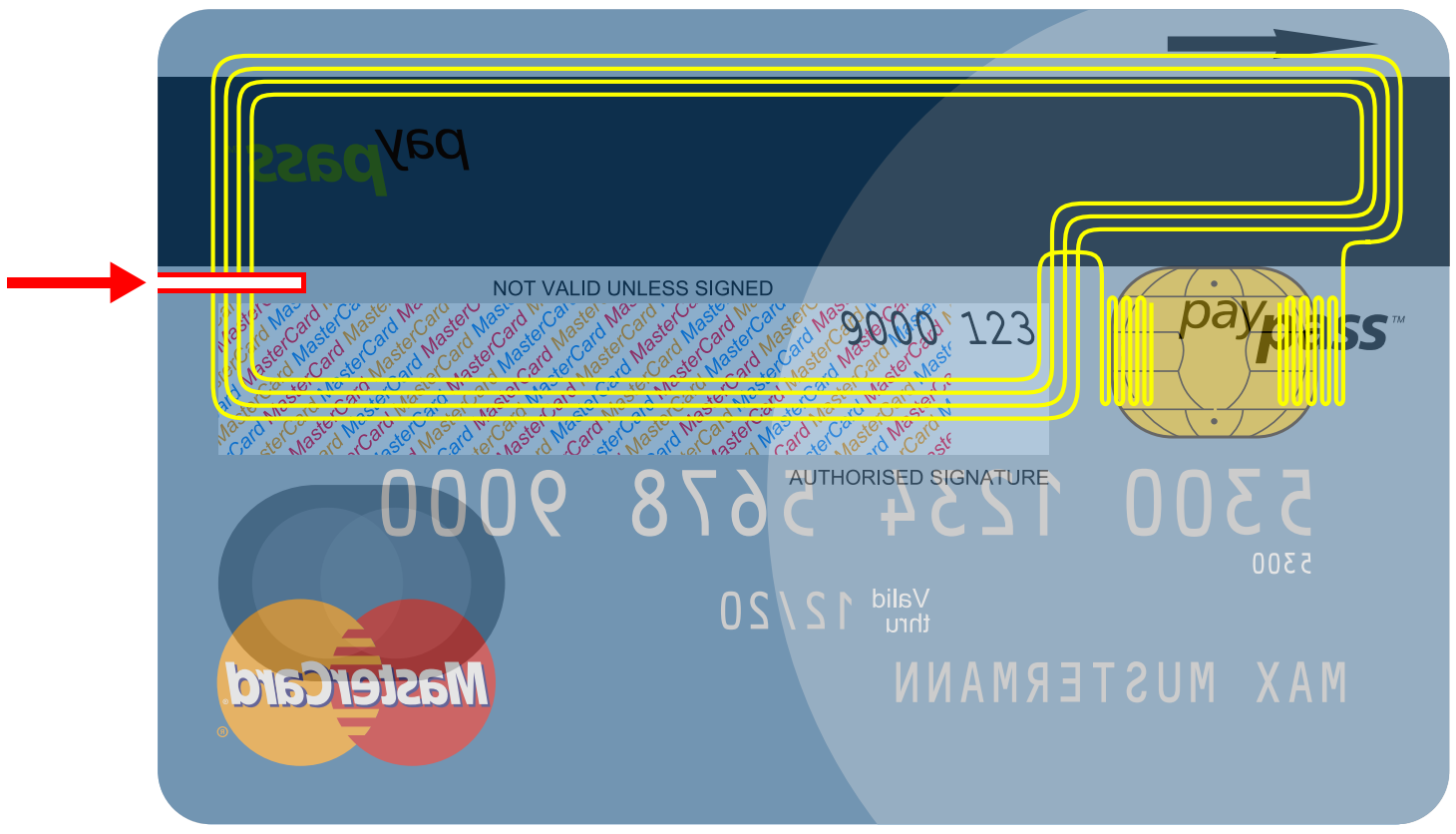

(a) optimal position (for our cards)

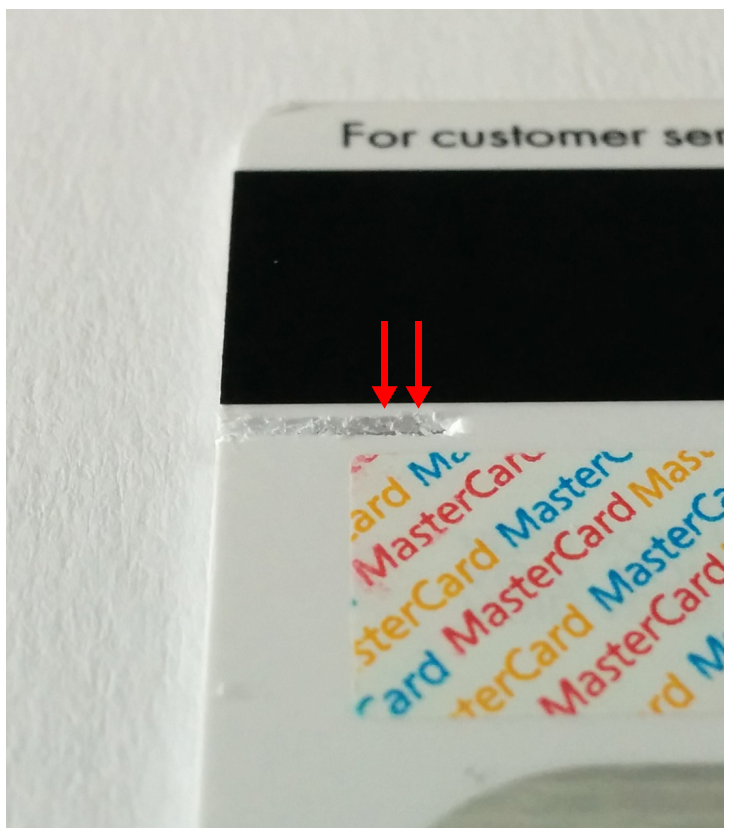

(b) 2 windings cut

Figure 14: Optimal position to cut the antenna loops

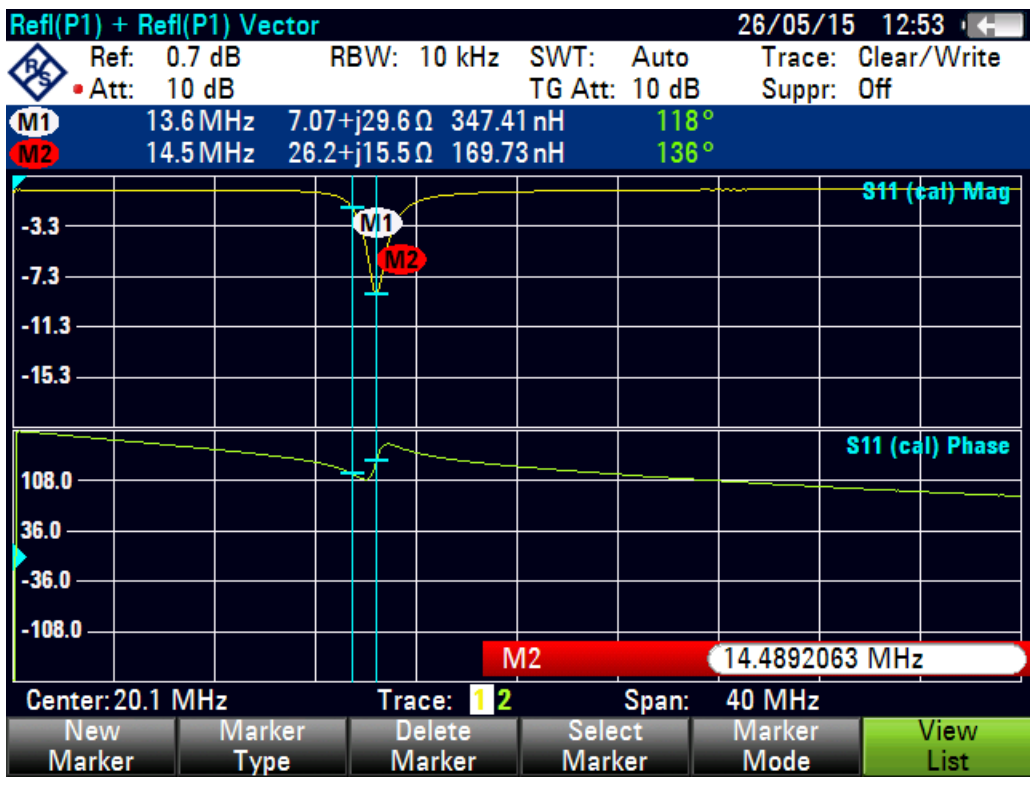

(a) all windings intact

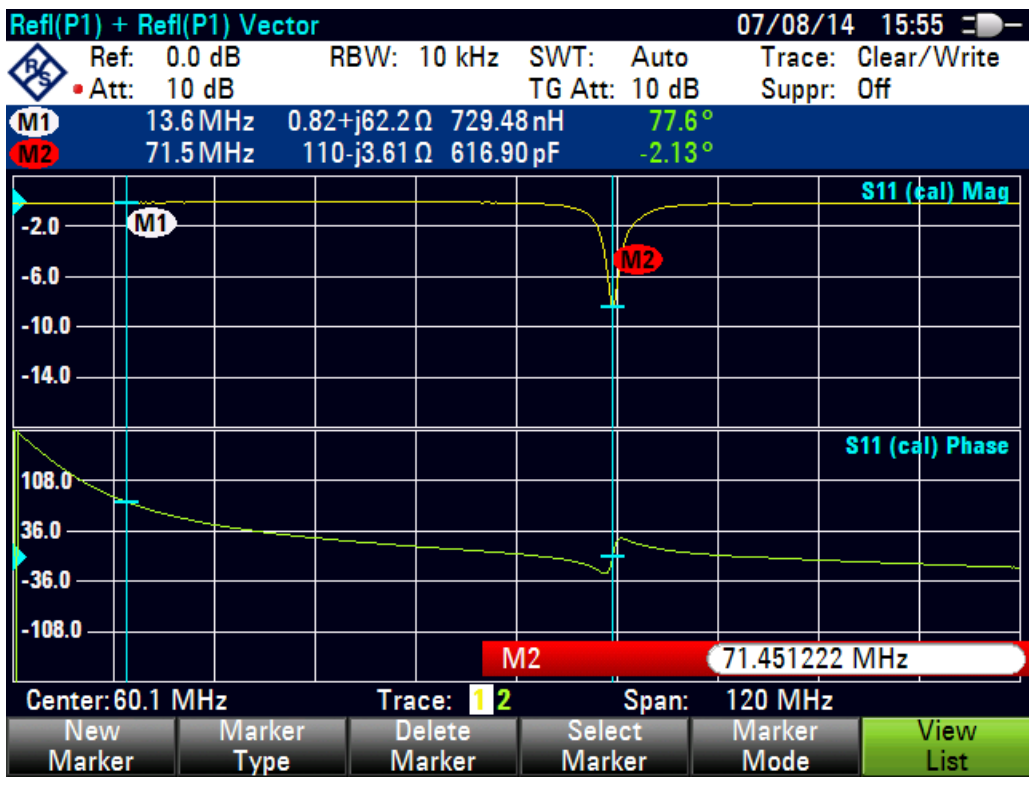

(b) 1 winding cut

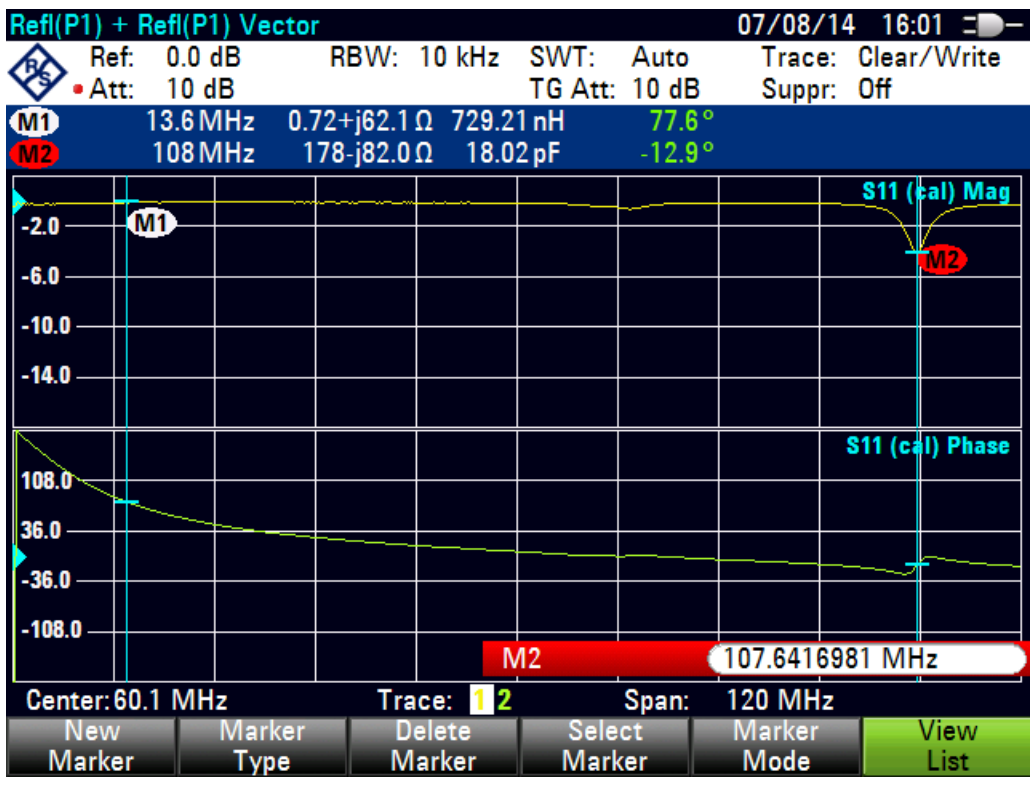

(c) 2 windings cut

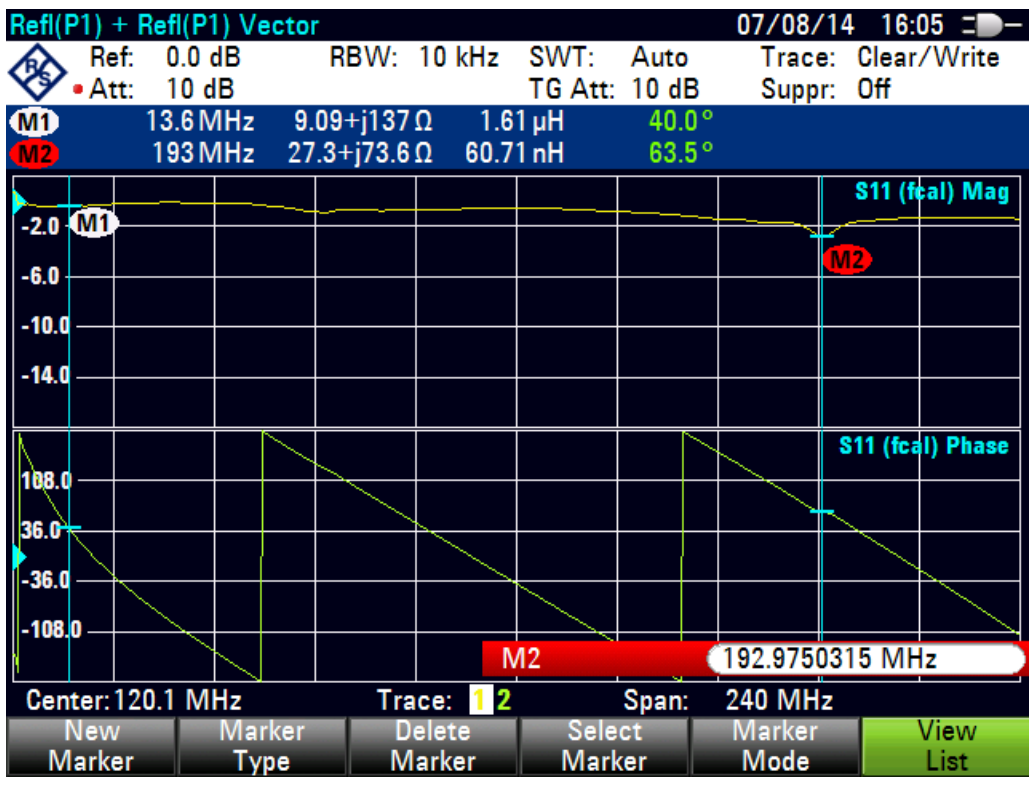

(d) 3 windings cut

Figure 15: Influence of cutting the antenna windings on the resonant frequency

on the resonant frequency and on the readability with smartcard reader hardware. Fig. 15 shows the change of the resonant frequency after cutting each wire:

1. Before cutting, the resonant frequency is around 15 MHz. The card can be read with an HID OMNIKEY 5321 smartcard reader (both through the contact interface and the contactless interface), as well as with an NFC-enabled Android smartphone (tested with Samsung Nexus S and LG Nexus 5).
2. After cutting the first winding, the resonant frequency increased to about 71 MHz. We were no longer able to read the card through its contactless interface.
3. After cutting the second winding, the resonant frequency increased to about 108 MHz.
4. After cutting the third winding, the resonant frequency increased to about 193 MHz.
5. After cutting the last winding, we could no longer observe any peaks in the magnitude of the reflection coefficient (S11).

In all cases, we could still successfully access the card through its contact interface.

In our tests, the contactless smartcard reader and the NFC-enabled smartphones could no longer communicate with the smartcard over the contactless interface after cutting the first winding. Therefore, it seems sufficient to cut the antenna in order to prevent, for instance, usage of a card for PIN-less micropayments at regular contactless payment terminals. However, this empirical test result cannot be generalized to all reader devices. An attacker with a specially crafted reader device, might still be capable of accessing the card over its contactless interface. Moreover, an attacker with unlimited physical access to the card (e.g. in the case of a lost or stolen card) could try to repair (by soldering the cut wire endings) or replace the RFID antenna.

Further analysis would be necessary to evaluate if smartcards with cut antenna windings could still be accessed over their contactless interface with a specially crafted reader device hardware. Cutting the antenna wire essentially adds a series capacitance to the antenna loop. This series capacitance detunes the smartcard antenna circuit to a higher resonant frequency, as can be seen in Fig. 15. Further research would need to analyze/simulate if sufficient power (as well as the data signal) can be transfered to the card and if responses by the card would still be detectable at the reader side after adding such a series capacitance.

## 4.2 Newer Antenna Concepts and their Possible Consequences

A new antenna technology for dual interface smartcards makes it possible to integrate an antenna into the card body that does not need wired contacts to the chip module (cf. [3, 7, 8]). Instead, the antenna in the card body has a few additional turns around the area where the chip module is embedded (see Fig. 16(c)). This card body antenna inductively couples into a tiny loop antenna that is directly integrated into the chip module (see Fig. 16(b)). This simplifies the card production process as the antenna does not need to be attached (e.g. glued, welded or soldered) to the chip module.

Unfortunately, this also means that the antenna of the chip module remains intact when the card antenna is cut through. Hence, the contactless interface of cards based on this antenna concept will likely remain accessible (at least with special reader hardware) even after the above mentioned method for disabling the contactless interface by cutting through the card antenna loops was applied.

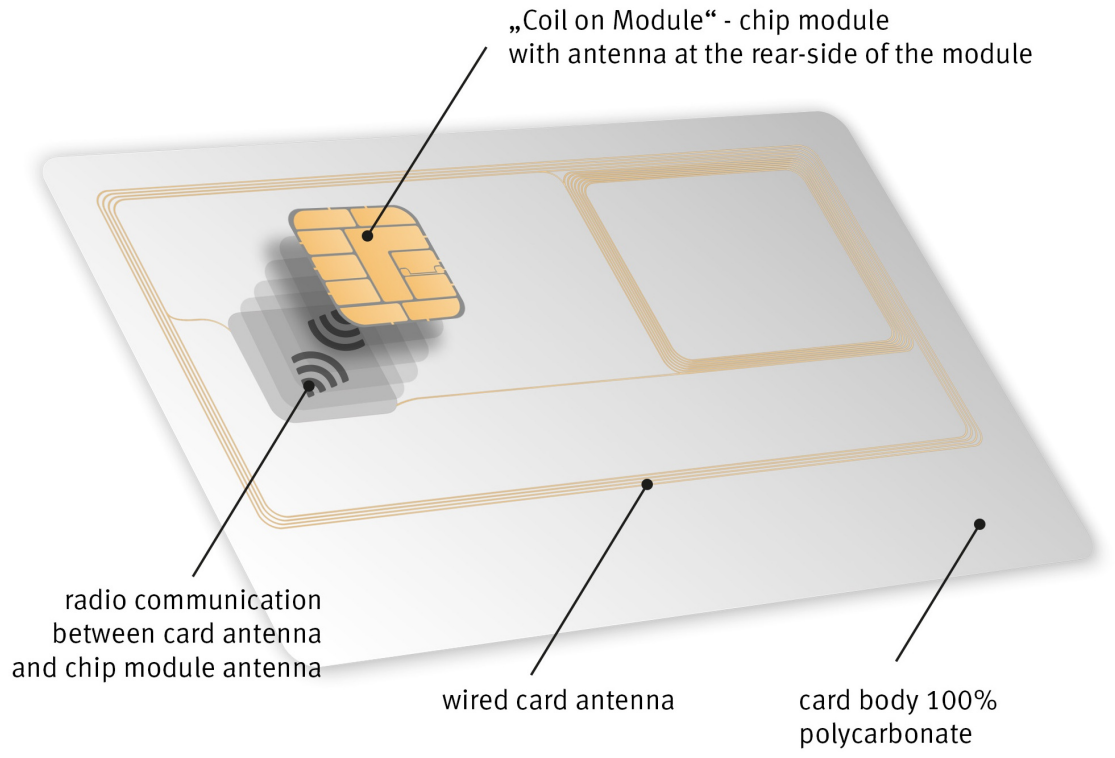

(a) card with inductively coupled chip module [9]

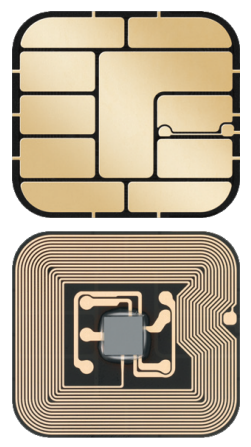

(b) chip module with integrated antenna [8]

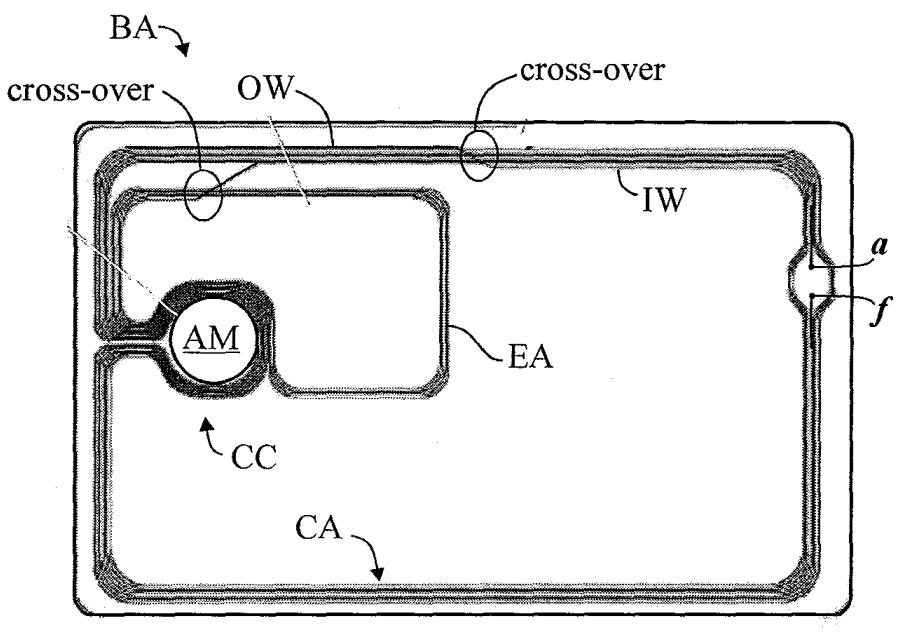

(c) card antenna [3]

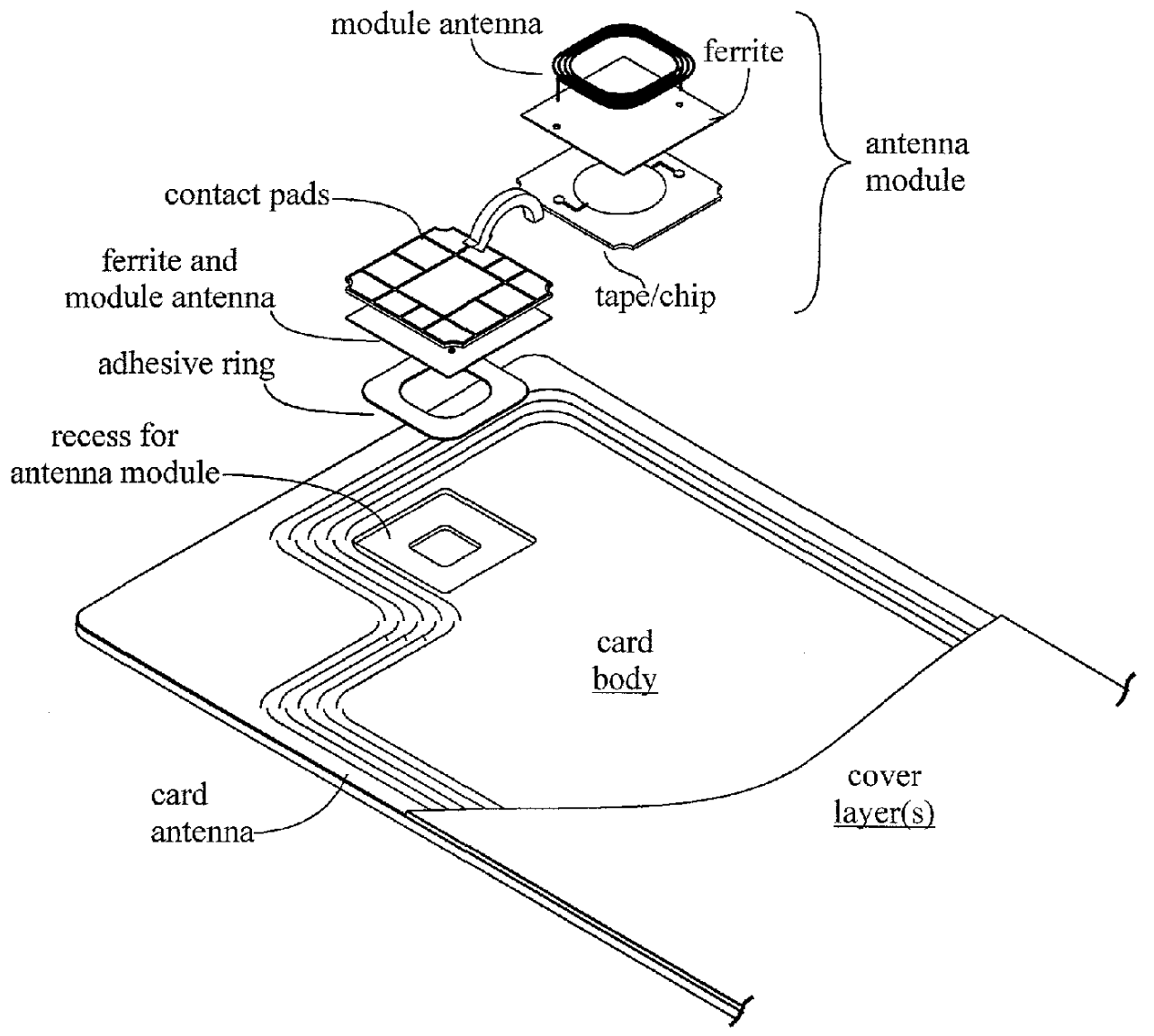

(d) stacked layers of chip module and embedding in card [3]

Figure 16: Card and chip module with separate, inductively coupled antennas

# 5. Smartcards with Switchable Contactless Interface

There are a number of patents and products related to smartcards with switchable contactless interface. For instance, US Pat. 10/334,572 [16], filed in 2002, introduces various concepts of mechanically switching the antenna loop of an RFID contactless card. The switch (see Fig. 17) is embedded into the card body and can be activated by pressing it or by using an activator token.

Similarly, US Pat. 11/503,197 [17] describes how the antenna of an RFID contactless card could be switched with a pressure-sensitive capacitive switch embedded into the card (see Fig. 18). As an example, Peratech[1] offers a technology to integrate switches into smartcards [12] and gives examples for chip cards with push-buttons as well as for inductive coupling cards with a switchable antenna [12].

US Pat. 13/701,883 [18] describes how bank cards with contact and/or contactless interface can be enhanced with a display and push-buttons. This concept uses a separate microcontroller that is capable of accessing the smartcard chip to control the display and the user inputs. Fig. 19 gives an overview of such a card. MasterCard uses such technology for their “Display Card” has an integrated LCD display and touch-sensitive buttons for generating one-time passwords [11].

While several concepts to disable the contactless interface of smartcards exist, most contactless and dual interface smartcards come without such mechanisms. Given the number of patents related to embedding switches into plastic cards and clipping the antenna of inductive-coupled contactless smartcards, we believe that it should be fairly easy to integrate such user-controlled switching capabilities into smartcards. We therefore further evaluated three approaches for implementing cards with a contactless interface that can be switched on and off by the user:

1. clipping the antenna with a switch,
2. short-circuiting the antenna close to the chip, and
3. switching the contactless interface through software on the smartcard chip.

## 5.1 Concept 1: Clipped Antenna

We found several patents that show how a push-button can be integrated into the antenna of a contactless smartcard in order to clip the antenna wire unless the user presses the button. The principle is rather simple (Fig. 20): A switch (normally open) is added in series to the antenna. When the user wants to give access to the contactless interface, the switch is closed and connects the antenna with the smartcard chip. So this is essentially a reversible variant of what we did by cutting the antenna windings in section 4.1.

[1] http://www.peratech.com/

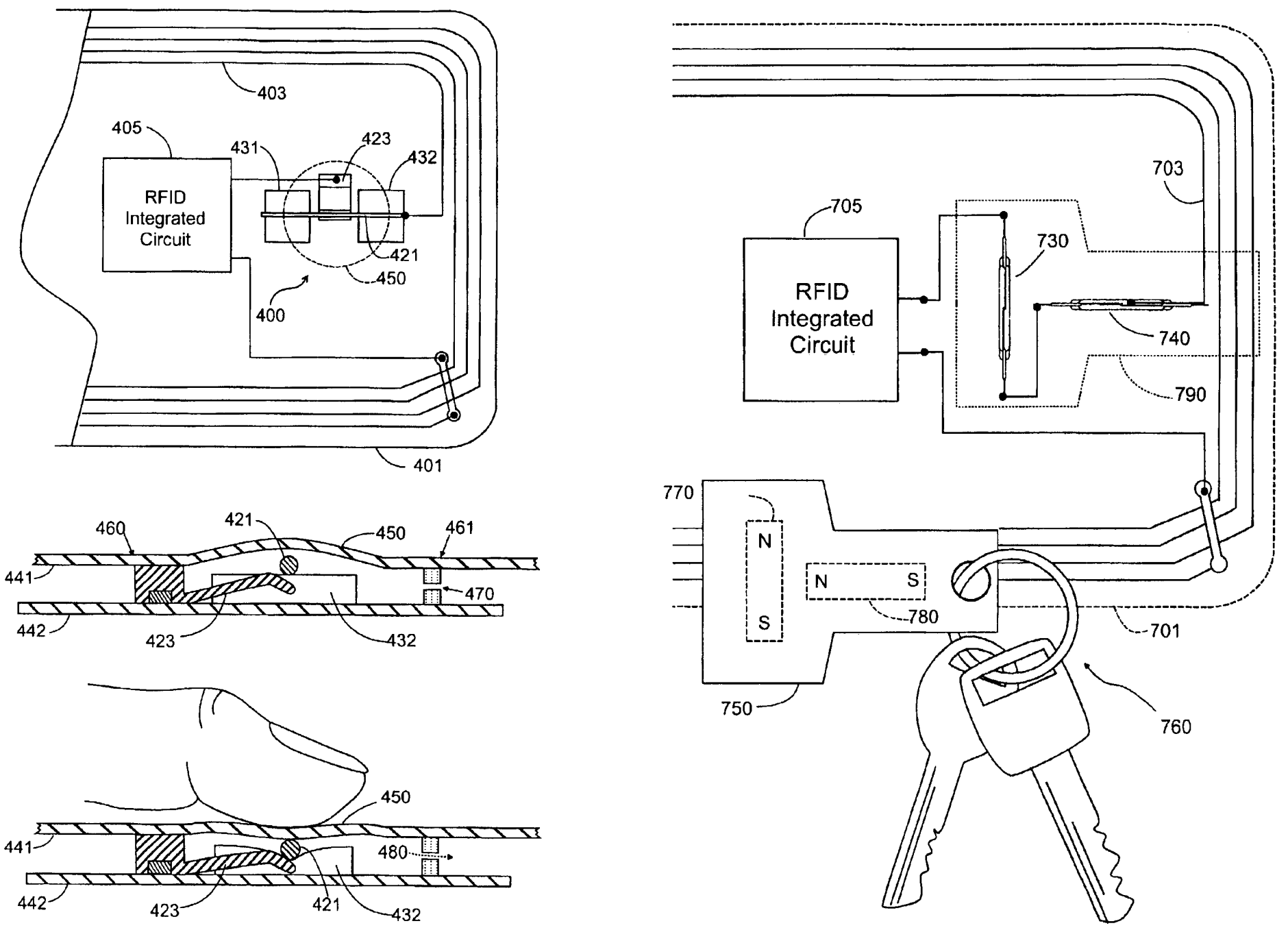

(a) mechanical button

(b) reed contact with magnetic activator key fob

Figure 17: Different approaches for mechanical switches embedded into contactless RFID cards [16]

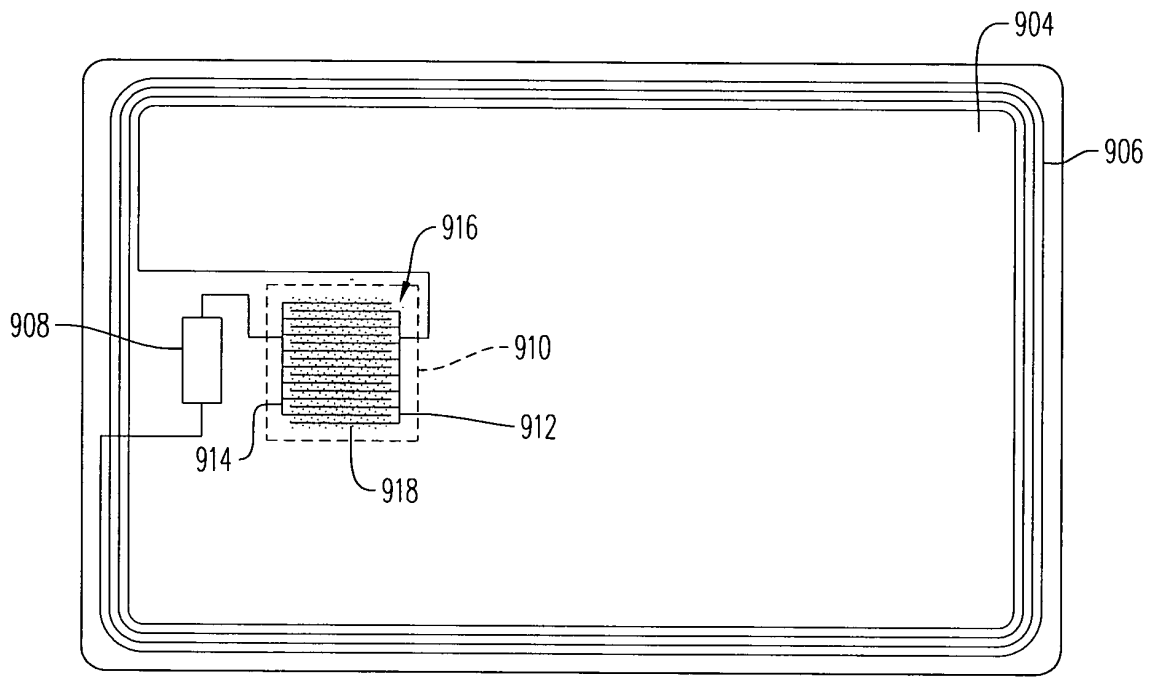

Figure 18: Pressure sensitive switch embedded into contactless RFID card [17]

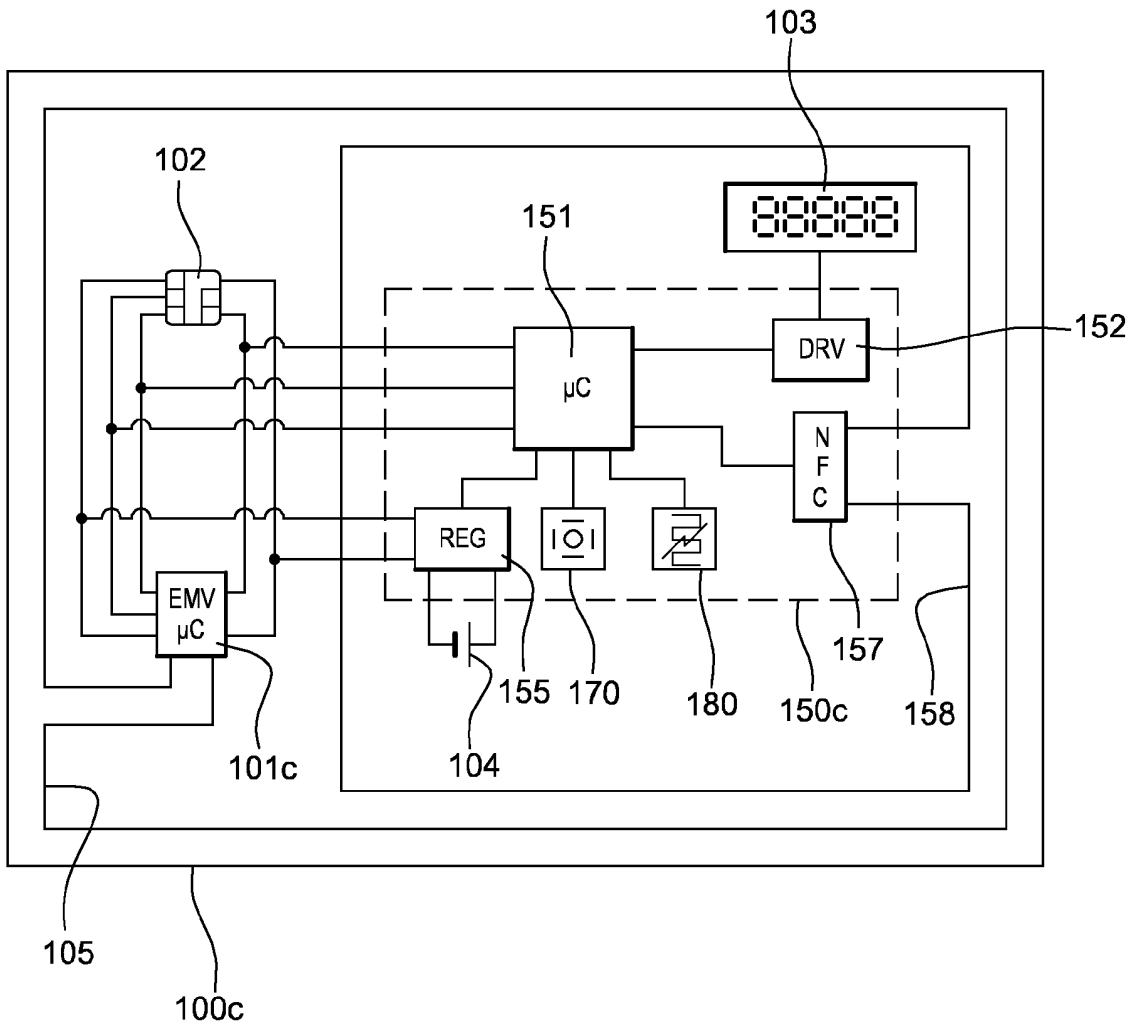

Figure 19: Smartcard enhanced with input and output capabilities [18]

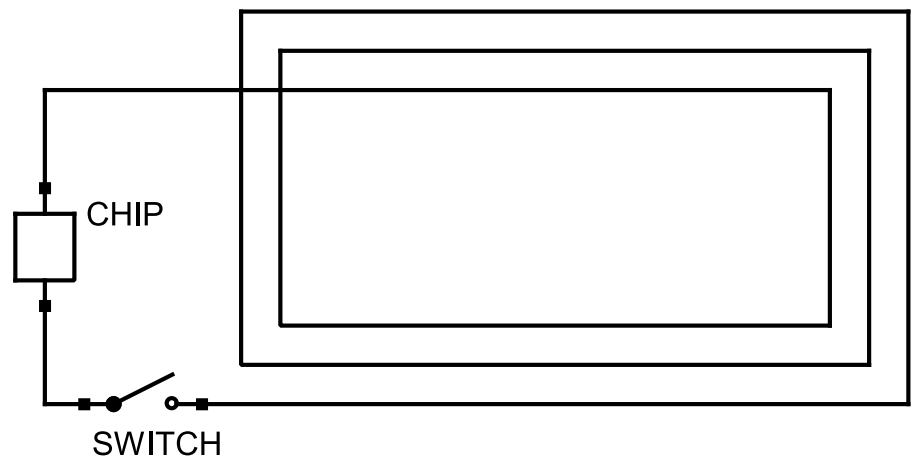

Figure 20: Clipping the antenna with a switch

While the idea seems simple, we did not find any smartcard products that actually make use of such technology. Therefore, we created a few proof of concept prototypes to assess the difficulties in building such a card based on existing smartcard chips. We used chips that we extracted (cf. section 2) from existing MIFARE Classic cards and EMV payment cards and used enameled copper wire to shape antennas similar to those of the plastic cards where we extracted the chips from.

### 5.1.1 MIFARE Classic

For our first prototype (Fig. 21), we prepared an ID-1 sized piece of cardboard and mounted the antenna along the edge of the cardboard. We placed two rectangular pads made of adhesive copper foil into one corner of the prototype. We soldered one end of the antenna directly to the antenna contact pad on the chip module and the other end to one of the copper pads. We then connected the second copper pad to the second antenna contact pad of the chip module. The antenna circuit can now be closed by connecting the two copper pads with a piece of metal (low resistance) or by simply placing a finger on top of them (high resistance). Moreover, we attached a pair of additional wires to the antenna contact pads to allow for further analysis of the signal at the antenna pads.

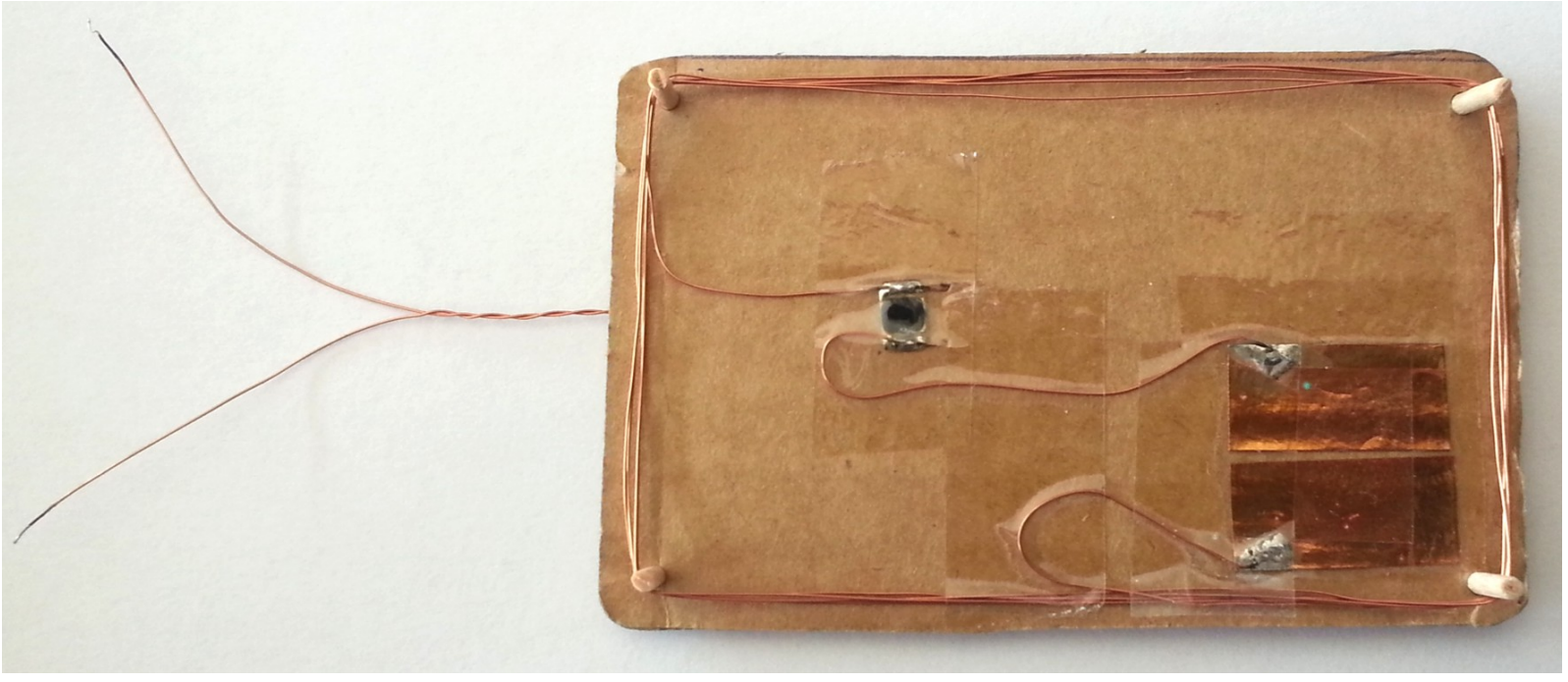

Figure 21: Prototype 1: MIFARE Classic on cardboard

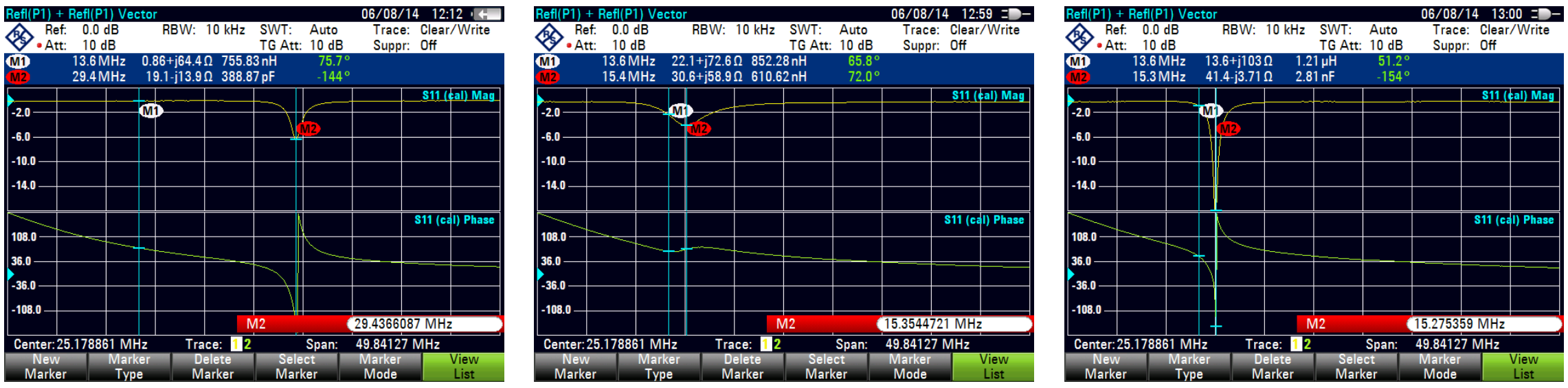

(a) copper pads unconnected

(b) copper pads connected with finger

(c) copper pads connected with metal bridge

Figure 22: Prototype 1: Measurement of the reflection coefficient

We found that our prototype could no longer be accessed using an HID OMNIKEY 5321 smartcard reader, as well as an NFC-enabled Android smartphone (tested with Samsung Nexus S and LG Nexus 5). After connecting the copper contact pads with either a metal bridge or a finger, the MIFARE Classic chip could be accessed using both, the contactless smartcard reader as well as the Android smartphones[2]. Fig. 22 shows that the switch detunes the resonant circuit of the tag. As discussed in section 4.1, further analysis would be necessary to determine if a specially crafted reader could be used to access the card despite the contact pads being unconnected.

As our first prototype worked quite well, we created a second prototype (Fig. 23) using a MIFARE Classic chip. This second prototype should demonstrate that it is possible to integrate the switching concept of prototype 1 into a laminated plastic card. We therefore took a chip including the antenna that we previously extracted from a plastic MIFARE Classic card. We then placed the antenna and the chip onto an ID-1 sized piece of adhesive plastic foil. We cut the antenna wire and soldered the two open ends to copper pads that we placed in one corner of the plastic foil. We then placed the piece of plastic foil in a thermal laminating film pouch. Before thermal lamination, we cut a square window in one side of the pouch where the upper side of the copper pads was located. Finally, we used a flat iron to thermally laminate the pouch around the inner plastic foil carrying our antenna.

[2]The LG Nexus 5 could only display the anti-collision identifier (UID) and the card type as the chipset of that device does not support the proprietary MIFARE Classic protocol.

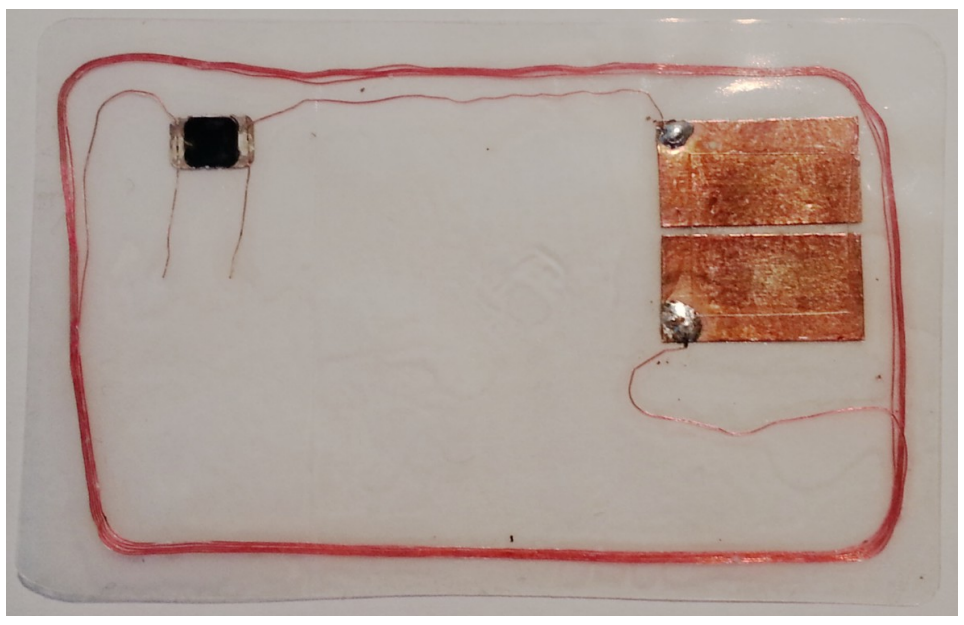

Figure 23: Prototype 2: laminated MIFARE Classic card

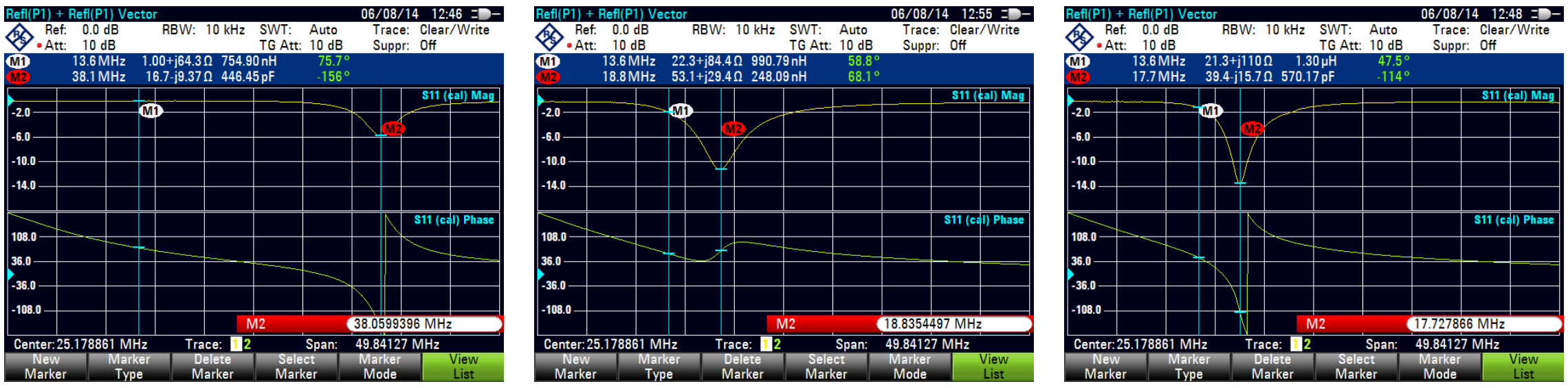

(a) copper pads unconnected

(b) copper pads connected with finger

(c) copper pads connected with metal bridge

Figure 24: Prototype 2: Measurement of the reflection coefficient

Tests showed that this card works equally well as our first prototype. Fig. 24 shows measurements of the resonant circuit of the card.

### 5.1.2 Dual Interface Processor Smartcard

We prepared a third prototype on an ID-1 sized piece of cardboard analogous to prototype 1. We added an antenna to the upper half of the “card” with similar dimensions as those of the card where we extracted the chip from. We placed two rectangular pads made of adhesive copper foil into one corner of the prototype and soldered those pads between one end of the antenna and the smartcard chip. Moreover, we attached a pair of additional wires to the antenna contact pads to allow for further analysis of the signal at the antenna pads. Fig. 25 shows a modified version of this prototype where the antenna is no longer routed through the copper pads and where two additional wires were soldered to pads of the chip module.

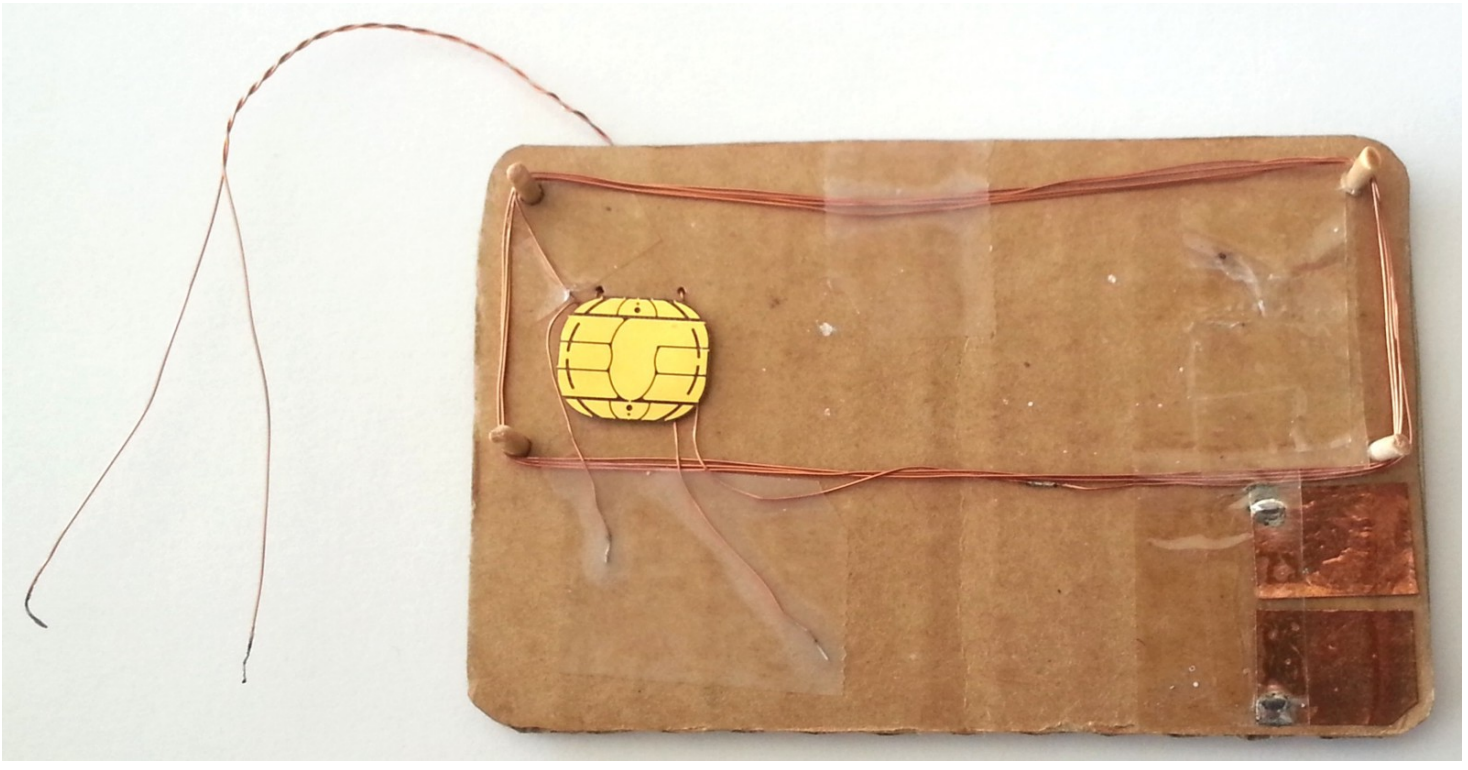

Figure 25: Prototype 3: Dual interface smartcard chip on cardboard

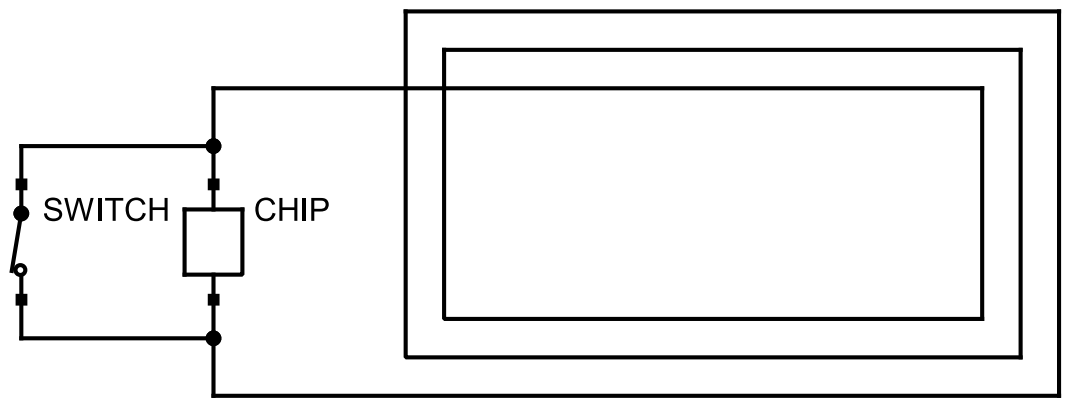

Figure 26: Short-cicuiting the antenna with a switch

We found that our prototype could no longer be accessed using an HID OMNIKEY 5321 smartcard reader, as well as an NFC-enabled Android smartphone (tested with Samsung Nexus S and LG Nexus 5). After connecting the copper contact pads with a metal bridge, the chip could be accessed using both, the contactless smartcard reader as well as the Android smartphones. Unfortunately, we were unable to get this prototype to work by putting a finger onto the copper pads. As the prototype worked using a metal bridge, we can assume that this failure was due to the high resistance through the human finger that was added in series to the antenna.

## 5.2 Concept 2: Short-Cicuited Antenna

As an alternative to breaking and closing the antenna loop, we also analyzed if it would be possible to short-circuit the antenna loop by directly connecting the antenna pads close to the chip module, hence, by placing a switch in parallel to the antenna (see Fig. 26). In this scenario, the switch would normally be closed and would, therefore, short-circuit the signal at the antenna pads. When the user wants to give access to the contactless interface, the switch is opened in order to allow the signal from the antenna to pass through to the smartcard chip.

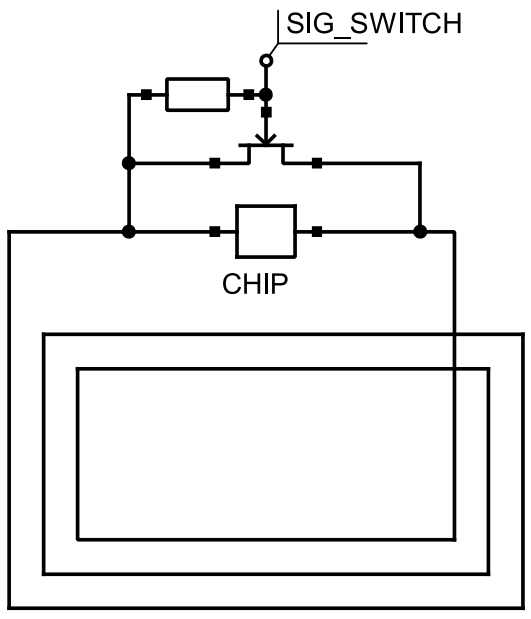

Figure 27: Short-cicuiting the antenna with a semiconductor switch

We tested this approach with our prototypes 1 and 3 by directly connecting the two copper contact pads to each other and by using the additional pair of wires that we attached to the antenna contacts to short-circuit the antenna. This method worked with both prototypes and they could no longer be accessed while the antenna contacts were directly connected to each other.

The disadvantage of this approach is that it requires a switch that is normally closed and has a low resistance when closed. Therefore, copper contact pads as in concept 1 are infeasible. Instead, a mechanical switch would be necessary. Alternatively, it might be possible to use a semiconductor switch (e.g. JFET) as shown in Fig. 27. However, this is beyond the scope of our evaluation and may be subject to future research.

## 5.3 Concept 3: On-Chip Switching of the Contactless Interface

Many typical dual-interface smartcards (specifically if compliant to Global Platform Card Specification Amendment C [5]) allow to enable and disable the contactless interface on a per-card basis and on a per-application (Java Card applet) basis. Moreover, applications can identify the source interface of commands and can employ their own interface-based policies for each command. Hence, such smartcards are capable of disabling access to functionality over the contactless interface in software. These capabilities could be used to create an applet that manages contactless activation and provides an ISO/IEC 7816-4 APDU-based interface to enable/disable contactless access.

### 5.3.1 Using Display Cards

“Display cards”, as, for instance, deployed by MasterCard [11], add a user interface to smartcards. With such cards, a separate microcontroller takes user inputs, generates user outputs, and interacts with the smartcard chip (cf. Fig. 19). Such cards could be used to enable and disable the contactless interface of a smartcard based on user inputs. Display cards as described in US Pat. 13/701,883 [18] typically use the ISO/IEC 7816 contact smartcard interface to interact with the smartcard chip. For example, pressing a button could trigger the microcontroller to send an APDU command to the management applet that instructs it to disable the contactless interface (possibly only for certain applications). Another button could trigger an APDU command that instructs the smartcard chip to (re-)enable its contactless interface.

#### 5.3.2 Using NFC-enabled Mobile Devices

Display cards require additional circuitry to be embedded into smartcards and consequently increase production costs. NFC-enabled smartphones could be used as a cheap alternative to interact with the management applet on the card. An app on the mobile device could provide a user interface, that lets the user select, which applications should be accessible over the contactless interface.

#### 5.3.3 Security Considerations for an Interface Management Applet

If contactless activation/deactivation is possible over the regular smartcard interfaces (ISO/IEC 7816 contact interface or ISO/IEC 14443 contactless interface), an attacker may be able to activate contactless access without user consent. For instance, if activation and deactivation is performed with simple enable and disable commands that are not protected from unauthorized access, an attacker might be able to issue those commands to activate interface access.

This is particularly problematic if the management applet is accessible over the contactless interface (e.g. if contactless interface activation is possible with an NFC-enabled mobile device). In that case, an attacker could simply send the activation command prior to accessing functionality that was supposedly protected by deactivation. As a countermeasure, the management applet could require authentication based on a user-supplied PIN code or password before accepting activation and deactivation commands. Such a PIN code or password could be easily provided by the user through the management app on the mobile device.

In the case of a display card, the management applet would typically be accessible only through the contact interface (as that is used by the microcontroller to access the smartcard chip). Hence, an attacker could not simply send an activation command over the contactless interface before accessing the protected applications over that exact same interface. Nevertheless, malicious contact smartcard terminals, could try to enable the contactless interface by automatically sending an activation command whenever a card is inserted. Therefore, even in that case it would make sense to protect the management applet with some form of authentication (cf. Google's flawed PIN code authentication in early versions of Google Wallet [15]).

A display card could even provide the necessary user input/output capabilities to request a PIN code from the user.

Besides protection against attackers without physical access to a card, requiring a user-supplied PIN code or password in order to activate certain functionality of a smartcard could even protect the card against attackers with physical access, an additional level of security that a simple push-button could not provide. However, it is questionable if users would accept the additional step of typing a PIN/password prior to using their card. After all, contactless transactions are often promoted as fast alternative to contact-based use of smartcards.

### 5.3.4 Smartcard Chips with Dedicated Switching Input

Another alternative would be to directly integrate a circuitry for switching the contactless interface based on user inputs into the smartcard microchip. A dedicated pin of the smartcard chip (accessible through the chip module) could be connected to a push-button or a capacitive input pad. Based on the sensed state of this switching input, the whole contactless interface could be enabled or disabled. This would provide a more cost-effective, reliable, and simple alternative to using a display card or an additional NFC-enabled mobile device. However, this would not protect against attackers with physical access to the card.

# 6. Summary

We showed how contactless and dual interface smartcards are constructed and where their antenna is located. We demonstrated simple and reliable methods to render the contactless interface of smartcards (permanently) unusable while maintaining all other functionality. Moreover, we presented and evaluated various concepts for making the contactless interface switchable by the end-user in order to improve the (perceived) security of contactless smartcards.